\documentclass[aps,prx,10pt,amssymb,amsmath,superscriptaddress,tightenlines,twocolumn,notitlepage,
] {revtex4-2}
\usepackage{graphicx}
\usepackage{amsmath}
\usepackage{amssymb}
\usepackage{bm}
\usepackage{color}
\usepackage{hyperref}
\usepackage{tabularx}
\usepackage{multirow}
\usepackage{booktabs}
\usepackage{braket}
\usepackage{epstopdf}

\begin{document}

\title{Universal giant spin Hall effect in moir\'e metal}

\author{Ning Mao}
\affiliation{Max Planck Institute for Chemical Physics of Solids, 01187, Dresden, Germany}

\author{Cheng Xu}
\affiliation{Department of Physics and Astronomy, University of Tennessee, Knoxville, TN 37996, USA}
\affiliation{Department of Physics, Tsinghua University, Beijing 100084, China}

\author{Ting Bao}
\affiliation{Department of Physics and Astronomy, University of Tennessee, Knoxville, TN 37996, USA}
\affiliation{Department of Physics, Tsinghua University, Beijing 100084, China}

\author{Nikolai Peshcherenko}
\affiliation{Max Planck Institute for Chemical Physics of Solids, 01187, Dresden, Germany}

\author{Claudia Felser}
\affiliation{Max Planck Institute for Chemical Physics of Solids, 01187, Dresden, Germany}

\author{Yang Zhang}
\affiliation{Department of Physics and Astronomy, University of Tennessee, Knoxville, TN 37996, USA}
\affiliation{Min H. Kao Department of Electrical Engineering and Computer Science, University of Tennessee, Knoxville, Tennessee 37996, USA}

\begin{abstract}
While moir\'e phenomena have been extensively studied in low-carrier-density systems such as graphene and semiconductors, their implications for metallic systems with large Fermi surfaces remain largely unexplored. Using GPU-accelerated large-scale \textit{ab-initio} quantum transport simulations, we investigate spin transport in two distinct platforms: twisted bilayer MoTe$_2$ (semiconductor, from lightly to heavily doping) and Nb$X_2$ ($X$ = S, Se; metals). In twisted MoTe$_2$, the spin Hall conductivity (SHC) evolves from $4\tfrac{e}{4\pi}$ at $5.09^\circ$ to $10\tfrac{e}{4\pi}$ at $1.89^\circ$, driven by the emergence of multiple isolated Chern bands. Remarkably, in heavily doped metallic regimes—without isolated Chern bands—we observe a universal amplification of the spin Hall effect from Fermi surface reconstruction under long-wavelength potential, with the peak SHC tripling from $6\tfrac{e}{4\pi}$ at $5.09^\circ$ to $17\tfrac{e}{4\pi}$ at $3.89^\circ$. For prototypical moir\'e metals like twisted Nb$X_2$, we identify a record SHC of $-17\tfrac{e}{4\pi}$ (-5200 $(\hbar / e)$$S/cm$ in 3D units), surpassing all known bulk materials. These results establish moir\'e engineering as a powerful strategy for enhancing spin-dependent transport, and advancing ab-initio methodologies to bridge atomic-scale precision with device-scale predictions in transport simulations.
\end{abstract}

\maketitle

When two layered materials are stacked with a mismatch, a moir\'e pattern emerges and generates a long-range periodic potential that significantly modifies the underlying electronic band structure~\cite{bistritzer2011moire,cao2018unconventional}. In momentum space, the moir\'e potential folds the original Brillouin zone into a smaller mini Brillouin zone, effectively transferring the original bands onto multiple intersecting minibands, leading to band inversion and rich topological phase diagrams~\cite{wu2019topological}. Moreover, in graphene or semiconductors, the Dirac pocket or band edge at each valley can be folded into a series of isolated topological flat bands (ITFBs)~\cite{kennes2021moire,mak2022semiconductor}. These topological flat bands serve as the building block for realizing the diverse topological and correlated phases including Mott insulators, superconductivity, and quantum anomalous Hall states~\cite{wang2024imaging,kang2020dirac,cai2023signatures,zeng2023thermodynamic,park2023observation,xu2023observation}. Recent experiments on twisted bilayer WSe$_2$ (t-WSe$_2$) and MoTe$_2$ (t-MoTe$_2$) reveal the emergence of double and triple quantum spin Hall (QSH) effects facilitated by these flat bands. t-WSe$_2$ exhibits a spin Hall conductivity (SHC) of $\tfrac{4e}{4\pi}$ at a twist angle near 3$^\circ$~\cite{kang2024observation}, whereas t-MoTe$_2$ has demonstrated an SHC of $\tfrac{6e}{4\pi}$ at a smaller angle of 2.1$^\circ$~\cite{kang2024evidence,xu2025multiple}. The giant spin current in moir\'e materials paves the way for future spintronic applications.

So far, most studies of the SHC in moir\'e systems have focused on topological flat bands, which reside within a narrow energy window—about 100 meV around the Fermi level—and thus limits current research to lightly doped moir\'e semiconductors~\cite{tao2024giant}. Naturally, one may ask whether it is possible to bypass these flat bands and seek an even higher SHC in the heavily doped metallic regimes of moir\'e semiconductors, or, more ambitiously, in moir\'e metals~\cite{zhang2024moire} with large Fermi surfaces. Under the influence of the moir\'e potential, these complex Fermi surfaces often host numerous band-inversion points, which can substantially enhance topology-related transport phenomena. Despite their importance and relative abundance, the transport properties of these moir\'e metallic systems remain largely unexplored. Two key factors make first-principle transport calculations challenging: the high computational cost of full diagonalization, which scales as $\mathcal{O}(N^3)$, and the requirement of an extremely dense k-mesh ($\sim 100 \times 100$ points) for convergence in the Kubo formula. Graphics processing units (GPUs) naturally address these challenges through massive parallelization~\cite{finkelstein2023fast}. By leveraging GPU parallelism, it becomes feasible to efficiently diagonalize large-scale Hamiltonians, enabling accurate exploration of transport properties across moir\'e materials.

In this work, we systematically investigate spin transport in t-MoTe\(_2\) from lightly doped to heavily doped metallic states, and extend our study to intrinsic moir\'e metals, twisted bilayer Nb\(X_2\) ($X$ = S, Se) (t-Nb$X_2$). In the lightly doped regime, we observe quantized SHC with peaks evolving from \(4\,\frac{e}{4\pi}\) at twist angles \(5.09^\circ\)--\(2.88^\circ\) to \(8\,\frac{e}{4\pi}\) for angles \(2.88^\circ\)--\(2.13^\circ\), and further increasing to \(10\,\frac{e}{4\pi}\) at \(2.00^\circ\) from the increasing number of isolated Chern bands. Strikingly, in the heavily doped metallic regime---despite the absence of isolated topological flat bands---we uncover a dramatic SHC amplification driven by moir\'e-induced Fermi surface reconstructions. At \(3.89^\circ\), the SHC peaks at \(17\,\frac{e}{4\pi}\), nearly fourfold larger than the maximum values in the multi-QSH states~\cite{kang2024evidence}. Remarkably, in t-Nb\(X_2\) at \(5.09^\circ\), we identify a colossal SHC of \(-17\,\frac{e}{4\pi}\), corresponding to \(-5200\ (\hbar / e)\mathrm{S/cm}\) in three-dimensional units. This value surpasses all known bulk spin Hall materials~\cite{nguyen2016spin,zhang2015role,guo2008intrinsic,zhang2021different} and is experimentally accessible without fine-tuning, as the peak resides precisely at the Fermi level.

\begin{figure}
\includegraphics[width=\columnwidth]{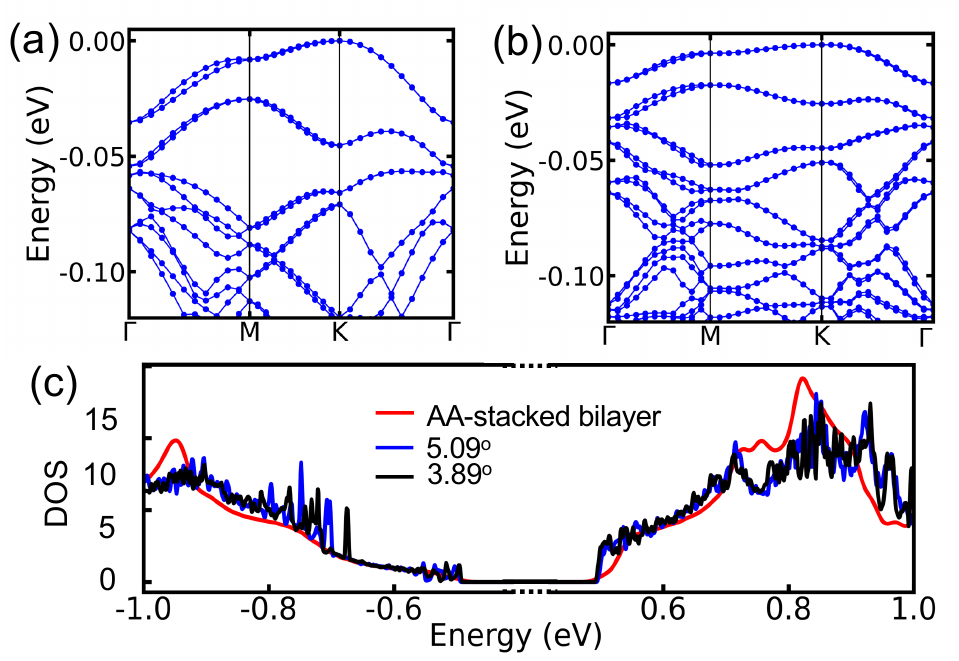}
\caption{Band structures of t-MoTe$_2$ with twist angles of (a) 5.09$^\circ$ and (b) 3.89$^\circ$. (c) Density of states within a 2 eV energy range, where the red, blue, and black lines correspond to the AA-stacked bilayer and t-MoTe$_2$ with twist angles of 5.09$^\circ$ and 3.89$^\circ$, respectively. All DOS values are normalized to that of the AA-stacked bilayer.
}\label{band}
\end{figure}

Starting with the two AA-stacked monolayers, we systematically construct t-MoTe$_2$ structures across a range of commensurate twist angles between 1.89$^\circ$ and 5.09$^\circ$. After obtaining the initial rigid moir\'e structures, we employ a transfer learning structure relaxation approach to perform large-scale lattice relaxation~\cite{mao2024transfer,bao2025transfer,zhang2024polarization,Jia2023}. We then feed the optimized structures into OpenMX for self-consistent calculations. Using the numerical pseudo-atomic orbitals (PAOs) as implemented in OpenMX, we focus on the electronic properties of the configurations~\cite{openmx_basis,openmx_largescale}. Here, PAOs are constructed from the non-orthogonal atomic orbitals:
\begin{equation}
 \left | \boldsymbol{r}, lm \right\rangle = R_l(r) Y_{lm}(\theta, \phi),
\end{equation}
where $R_l(r)$ is the radial function and $Y_{lm}(\theta, \phi)$ are spherical harmonics describing the angular part of the function. For simplicity, we use $\alpha$ to represent the combined index $lm$ in the following. We set the convergence criterion for the self-consistent field (SCF) calculations to a total energy difference smaller than 6 $\times$ 10$^{-5}$ Hartree between the last two SCF loops.

Upon convergence, we obtain the real-space Hamiltonian \(H_{\alpha\beta}(\mathbf{R})\), the overlap matrix \(S_{\alpha\beta}(\mathbf{R})\), and the Wannier center matrix elements \(A_{\alpha\beta}(\mathbf{R})\) as follows:
\begin{equation}
\begin{aligned}
H_{\alpha\beta}(\mathbf{R}) &= \langle 0\alpha \mid \hat{H} \mid \mathbf{R}\beta \rangle,\\
S_{\alpha\beta}(\mathbf{R}) &= \langle 0\alpha \mid \mathbf{R}\beta \rangle,\\
A_{\alpha\beta}(\mathbf{R}) &= \langle 0\alpha \mid \hat{r} \mid \mathbf{R}\beta \rangle,
\end{aligned}
\end{equation}
where \(\alpha\) and \(\beta\) label different orbitals, and \(\mathbf{R}\) is the lattice vector (i.e., the displacement from the reference cell at \(\mathbf{R}=0\) to the cell containing orbital \(\beta\)). Here, \(\hat{H}\), \(\hat{S}\), and \(\hat{r}\) denote the Hamiltonian, overlap, and position operators, respectively.
Applying a Fourier transform, we obtain the Hamiltonian and overlap matrices in reciprocal space:
\begin{equation}
\begin{aligned}
H_{\alpha \beta}(\mathbf{k}) &= \sum_{\mathbf{R}} e^{i \mathbf{k} \cdot \mathbf{R}} \,H_{\alpha \beta}(\mathbf{R}),\\
S_{\alpha \beta}(\mathbf{k}) &= \sum_{\mathbf{R}} e^{i \mathbf{k} \cdot \mathbf{R}} \,S_{\alpha \beta}(\mathbf{R}).
\end{aligned}
\end{equation}
We then solve the following generalized eigenvalue problem:
\begin{equation}
\bigl(H(\mathbf{k}) - E(\mathbf{k})\,S(\mathbf{k})\bigr)\,V(\mathbf{k}) = 0.
\end{equation}
Subsequently, we can use the coefficients $V(\mathbf{k})$ to write the Bloch wavefunction as
\begin{equation}
\left|\Psi_{\alpha,\mathbf{k}}\right\rangle
=\sum_{\,\beta}\,V_{\beta \alpha}(\mathbf{k})\,\left|\mathbf{k}\,\beta\right\rangle
= \sum_{\mathbf{R},\,\beta} e^{i\mathbf{k}\cdot \mathbf{R}}\,V_{\beta \alpha}(\mathbf{k})\,\left|\mathbf{R}\,\beta\right\rangle.
\end{equation}
Here, the coefficients \( V(\mathbf{k}) \) are not strictly orthogonal. Instead, they satisfy the generalized orthogonality condition:
\begin{equation}
V^\dagger(\mathbf{k}) S(\mathbf{k}) V(\mathbf{k}) = I,
\end{equation}
where \( I \) is the identity matrix.

Due to the high dimensionality of our matrices—specifically, a dimension of 45,132 at a twist angle of 3.89$^\circ$—it becomes computationally challenging to directly diagonalize the full Hamiltonian using conventional CPU methods. To overcome this limitation, we first apply Cholesky decomposition to transform the generalized eigenvalue problem into a standard eigenvalue problem:
\begin{equation}
C y = \lambda y, \quad C = L^{-1} H(\mathbf{k}) L^{-T}, \quad S(\mathbf{k}) = L L^{T},
\end{equation}
where $C$ and $L$ denote the transformed Hamiltonian matrix and Cholesky factor.
We then employ a two-stage dense matrix diagonalization technique (as detailed in Supplementary Material Section VII). Firstly, the dense matrix $C$ is converted into a banded form. Subsequently, we utilize the divide-and-conquer algorithm to obtain all eigenvalues and eigenvectors of this banded matrix. Finally, we recover the original eigenvectors using $x = L^{-T} y$. This approach is particularly suitable for GPU parallel architectures, as it allows simultaneous eigenvalue calculations for multiple smaller matrix blocks. By leveraging these algorithms, we efficiently compute the complete set of eigenvalues and eigenvectors. Figures~\ref{band}(a) and \ref{band}(b) illustrate the band structures near the Fermi level at twist angles of 5.09$^\circ$ and 3.89$^\circ$. At both twist angles, the two valence bands nearest the Fermi level can be approximately treated as isolated flat bands. Specifically, at a twist angle of 5.09$^\circ$, the two uppermost bands exhibit relatively large bandwidths of approximately 35 meV and 39 meV. At 3.89$^\circ$, due to the increased moir\'e superlattice periodicity and the resulting shrinkage of the Brillouin zone, the two topmost bands demonstrate reduced bandwidths of 17 meV and 14 meV.

To provide a more intuitive representation of the electronic state distribution within the Brillouin zone, we calculated the density of states (DOS) according to the equations in Supplementary Material Section III. As shown in Fig.~\ref{band}(c), the DOS exhibits two prominent peaks near the Fermi level, each corresponding to one of the two flat bands. Notably, the DOS peak at a twist angle of 3.89$^\circ$ is closer to the Fermi energy and displays a significantly higher amplitude compared to the one at 5.09$^\circ$. This indicates that the flat band at 3.89$^\circ$ is considerably flatter. When normalizing the DOS of t-MoTe$_2$ against that of the corresponding bilayer system, we find no significant global enhancement in the number of states.

\begin{figure}[t]
\includegraphics[width=1\columnwidth]{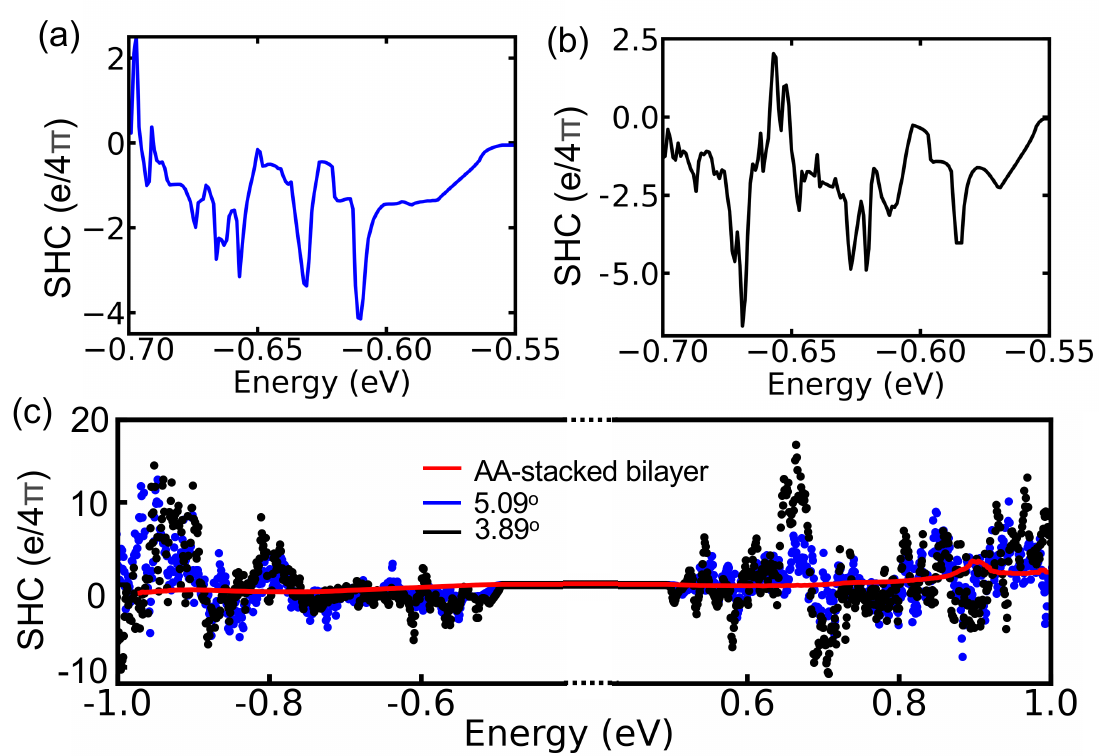}
\caption{
Zoomed-in views of the SHC over a 0.15 eV energy window for t-MoTe$_2$ with twist angles of (a) 5.09$^{\circ}$ and (b) 3.89$^{\circ}$, respectively. Quantized SHC plateaus emerge near the flat bands. (c) Broad energy-range comparison of the SHC over 2 eV for the untwisted bilayer and t-MoTe$_2$ at 5.09$^{\circ}$ and 3.89$^{\circ}$. The SHC in twisted structures is enhanced by several orders of magnitude compared to the untwisted bilayer, in stark contrast to the DOS shown in Fig. 1(c). All calculations employ a 100 $\times$ 100 k-point mesh within the Kubo formalism, using a broadening parameter of 10 meV to capture long-range SHC enhancements.
}\label{remote}
\end{figure}

We now discuss the topological properties of these isolated flat bands. In t-MoTe$_2$, the low-energy valence bands originate predominantly from the $\pm K$ valleys of the untwisted bilayer. Due to strong spin-orbit coupling (SOC) and the absence of inversion symmetry, the electron spins are locked along the $z$-axis, and opposite valleys exhibit opposite spin orientations. This behavior reflects an Ising-type SOC described by $H_\mathrm{SOC} = \lambda\,\tau_z \cdot \sigma_z,$ where $\tau_z$ and $\sigma_z$ act on the valley and spin spaces, respectively~\cite{gmitra2017proximity,bawden2016spin}. Consequently, there is no strong spin mixing in the low-energy valence band, allowing $S_z$ to be treated as a good quantum number. To separate the eigenvectors associated with different valleys for subsequent topological analysis, we perform a transformation on the degenerate bands,
\begin{equation}
|\tilde{\Psi}_{n,\mathbf{k}} \rangle = \sum_m U_{nm} (\mathbf{k})|\Psi_{m,\mathbf{k}}\rangle,
\end{equation}
where the transformation matrix $U(\mathbf{k})$ is given by
\begin{equation}
U =
\begin{pmatrix}
 \cos\theta & e^{-i\delta} \sin \theta \\
e^{i\delta} \sin \theta &  \cos \theta
\end{pmatrix},
\end{equation}
with $\theta$ and $\delta$ being the $\mathbf{k}$-dependent parameters.
To find the optimal parameter, we should evaluate the expectation values of the valley operator ($\hat{s}_z$) for these eigenvectors as~\cite{ezawa2013high,ezawa2012valley}
\begin{equation}
S^{z}_{mn}(\mathbf{k}) = \bigl\langle \Psi_{m,\mathbf{k}} \bigm| \hat{s}_z \bigm| \Psi_{n,\mathbf{k}} \bigr\rangle.
\end{equation}
Then, eigenvectors with $S^{z}_{mn}(\mathbf{k}) \approx +1$ or $-1$ can be associated with the $+K$ or $-K$ valley, respectively. Hence, half of the eigenvectors correspond to the $+K$ valley, while the other half correspond to the $-K$ valley. In the following topological discussion, we focus on the $+K$-valley.

Next, we compute the Chern band using the Fukui--Hatsugai--Suzuki method, where the Berry curvature within each plaquette is given by~\cite{fukui2005chern}
\begin{equation}
\begin{aligned}
&\Omega(\mathbf{k}) = \mathrm{Im} \ln \bigl( F(\mathbf{k},  k_x)
F(\mathbf{k}+ k_x,\,\mathbf{k}+ k_x+ k_y)  \\
&F(\mathbf{k}+ k_x+ k_y,\,\mathbf{k}+ k_y)
F(\mathbf{k}+ k_y,\,\mathbf{k})
\bigr).
\end{aligned}
\end{equation}
Here $F(\boldsymbol{k},\,\boldsymbol{k}+\delta\boldsymbol{k})$ is the overlap matrix, as defined in the Supplemental Material~\cite{SM}. The Chern number is then evaluated as the integral of the Berry curvature.
$C \;=\; \frac{1}{2\pi}\,\sum_{\boldsymbol{k}}\,\Omega(\boldsymbol{k}).$
Consequently, for twist angles between $5.09^\circ$ and $3.15^\circ$, the Chern number of the three flat bands remains $(1,1,-2)$, while at $2.88^\circ$ it becomes $(1,1,0)$. For twist angles between $2.65^\circ$ and $2.13^\circ$, four ITFBs with a Chern number of 1 are observed. At $2.00^\circ$ and $1.89^\circ$, five ITFBs with a Chern number of 1 show up. We also confirmed the Chern number through Wilson loop and edge states (as detailed in the Supplementary Material Sections IV and V)
~\cite{SM}.

Focusing on the ITFBs near the Fermi level, we proceed to investigate their transport properties using the Kubo formula. To ensure numerical convergence, a sufficiently dense k-mesh ($100 \times 100$) is employed for the moir\'e Brillouin zone. Increasing the grid size to $200 \times 200$ alters the SHC by less than $5\%$. Owing to the presence of three-fold rotation symmetry, two-fold rotation symmetry, and time-reversal symmetry in the t-MoTe$_2$ system, points in the Brillouin zone related by these symmetries share identical spin Berry curvature (as detailed in Supplementary Material VI). Exploiting these symmetries, we significantly simplify the k-point sampling from an original set of 10,000 $k$-points to 884 irreducible $k$-points. This substantial reduction considerably decreases computational costs, allowing the Kubo formula calculations to converge effectively with a relatively dense mesh. Here, the matrix elements of the momentum operator are expressed as~\cite{wang2019first,jin2021calculation,lee2018tight}
\begin{equation}
\begin{aligned}
&\langle \mathbf{k}\alpha | p| \mathbf{k}\beta \rangle =
\sum V_{\mathbf{k}\alpha}^*  V_{\mathbf{k}\beta}
\left( \frac{\partial H(\mathbf{k})}{\partial \mathbf{k}} - E_{\mathbf{k}\alpha} \frac{\partial S(\mathbf{k})}{\partial \mathbf{k}} \right) \\
&\quad + i \left(E_{\mathbf{k}\alpha} - E_{\mathbf{k}\beta}\right)
\sum V_{\mathbf{k}\alpha}^*  V_{\mathbf{k}\beta}
\sum_{\mathbf{R}} A(R) e^{i\mathbf{k} \cdot \mathbf{R}},
\end{aligned}
\end{equation}
The SHC is then evaluated as
\begin{equation}
\sigma_{xy}^{S} \;=\; e\,\hbar \int \frac{d^3k}{(2\pi)^3}\,\Omega^S(\mathbf{k}),
\end{equation}
where the spin Berry curvature $\Omega^S(\mathbf{k})$ is defined by
\begin{equation}
\Omega^S(\mathbf{k})=
-2\mathrm{Im}\sum_{\substack{ \alpha\neq \beta}}
\frac{
\langle \mathbf{k}\alpha \vert J_x^s(\mathbf{k}) \vert \mathbf{k}\beta \rangle
\langle \mathbf{k}\beta \vert p_y \vert \mathbf{k}\alpha \rangle
}{
\bigl(E_{\mathbf{k}\beta} - E_{\mathbf{k}\alpha}\bigr)^2
}.
\end{equation}
Here, the spin current operator along $x$ is given by
$
J_x^s(\mathbf{k})
\;=\;
\frac{\hbar}{2}\,\bigl\{\,s_z'(\mathbf{k}),\,p_x\bigr\},$ with $s_z'(\mathbf{k}) = S(\mathbf{k}) \ast s_z$, where $s_z$ is the $z$-component of the spin operator and $p_x$ (or $v_x$) denotes the corresponding momentum (or velocity) operator in the $x$-direction.

\begin{figure}[t]
\includegraphics[width=1\columnwidth]{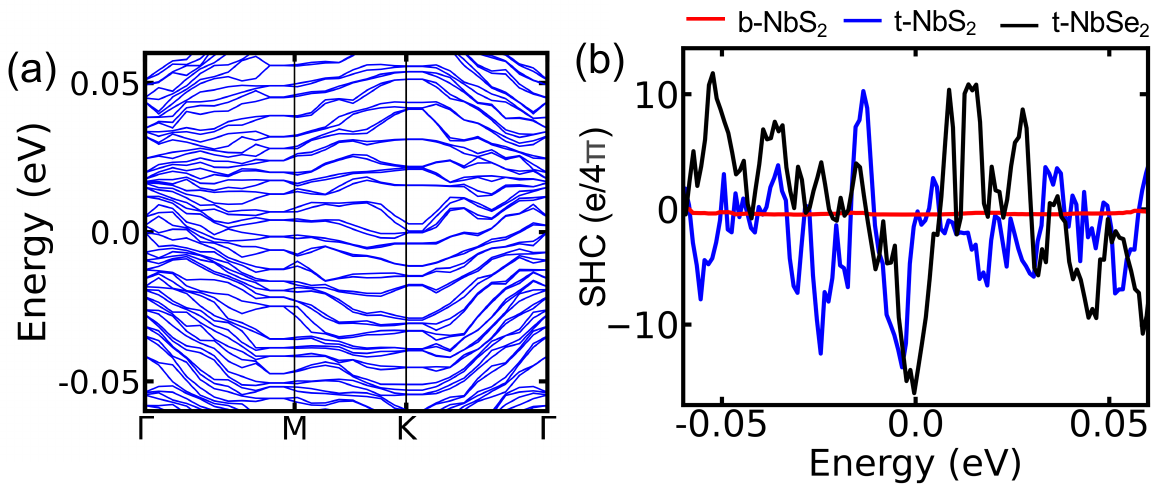}
\caption{
(a) Band structure of t-NbSe$_2$ with a twist angle of $5.09^\circ$.
(b) Variation of the SHC with energy, where the red, blue, and black curves correspond to bilayer NbS$_2$, t-NbS$_2$, and t-NbSe$_2$, respectively.
}\label{NbS}
\end{figure}

As shown in Fig.~\ref{remote}, the spin Hall effect displays universal amplification across doping regimes but reveals fundamentally distinct angular dependencies. In lightly doped regimes, quantized SHC plateaus emerge from isolated topological flat bands, solely dependent on the number of flat Chern bands and showing limited angular sensitivity. At $5.09^\circ$ twist angle, the maximum quantized SHC reaches $4\frac{e}{4\pi}$, increasing only to $10\frac{e}{4\pi}$ at $1.89^\circ$. Notably, between $5.09^\circ$ and $2.88^\circ$, the SHC remains fixed at $4\frac{e}{4\pi}$, demonstrating the inherent stability of ITFB-derived spin transport against angular variations.

In stark contrast, metallic regimes exhibit dramatically enhanced angular sensitivity despite lacking ITFBs. These systems feature gapless, massively entangled band structures that generate unexpectedly strong SHE contributions surpassing those from topological flat bands. The SHC increases nonlinearly with decreasing twist angle, reaching a maximum amplification of $17\frac{e}{4\pi}$ at $3.89^\circ$---nearly triple the $6\frac{e}{4\pi}$ value observed at $5.09^\circ$. As shown in Fig.~\ref{remote}(c), this enhancement peaks approximately 170 meV above the conduction band edge, where moir'e-induced band inversions maximize Berry curvature density. The metallic regime's combination of giant response and strong angle dependence suggests superior experimental viability compared to quantization-limited topological bands.

\begin{figure}[t]
\includegraphics[width=1\columnwidth]{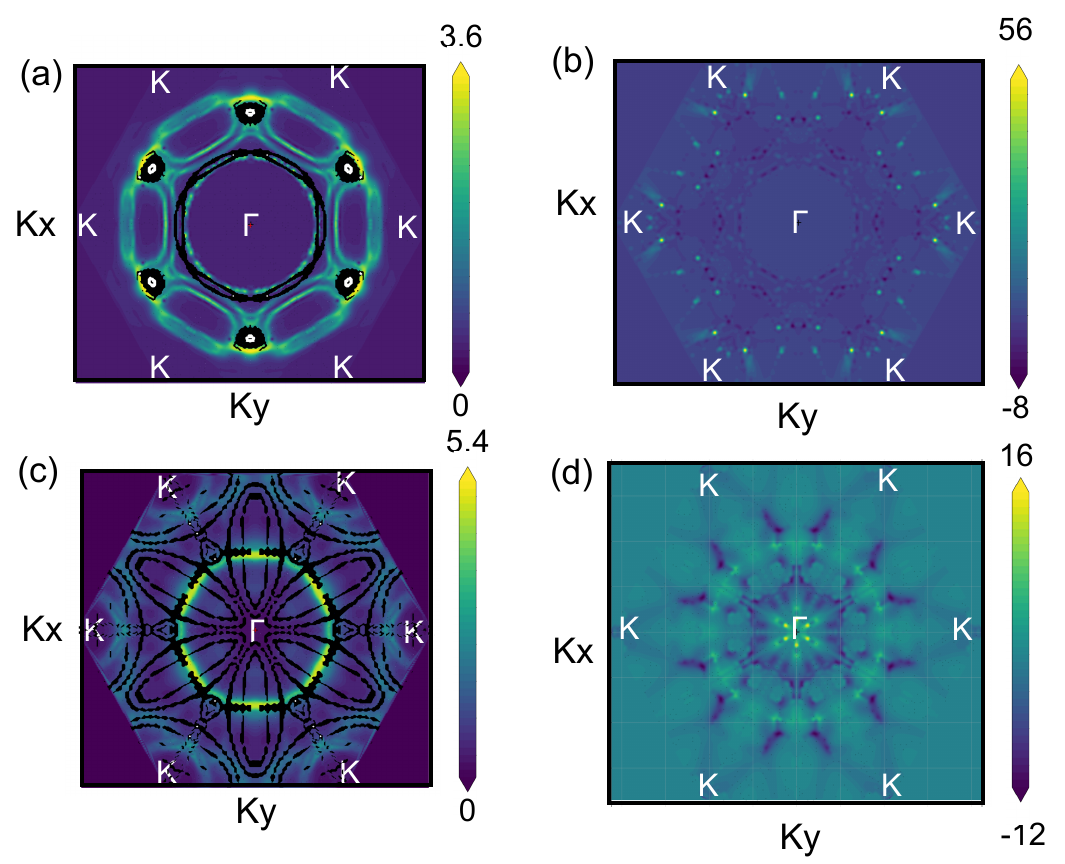}
\caption{
(a) Folded Fermi surface within the Brillouin zone of a monolayer, where the black line indicate the Fermi surface of monolayer MoTe$_2$.
(b) Spin Berry curvature distribution of t-MoTe$_2$.
In both panels, the Fermi energy is set to 170~meV above the conduction band minimum, where the SHC can increase to approximately 6$\frac{e}{4\pi}$.
(c) Folded Fermi surface within the Brillouin zone of a monolayer for NbSe$_2$. (d) Spin Berry curvature distributions of t-NbSe$_2$, where the Fermi energy is set at the charge neutrality point (zero energy).
}\label{sberry}
\end{figure}

Extending this concept to intrinsic moir\'e metal systems~\cite{zhang2024moire}, we note that niobium (Nb, $4d^4 5s^1$) has one fewer valence electron compared to molybdenum (Mo, $4d^5 5s^1$), thereby shifting its Fermi energy downward by approximately 1 eV. Consequently, Nb$X_2$ compounds can form large Fermi surfaces, covering about 48\% of the Brillouin zone, which, upon stacking, naturally give rise to a sequence of band inversions under long-wavelength moir\'e potential. We therefore construct a moir\'e superlattice by stacking two layers of 2H-NbS$_2$ or NbSe$_2$ with a twist angle of $5.09^\circ$. As shown in Fig.~\ref{NbS}(a), the twisted structure remains metallic and features a large and complex Fermi surface near the Fermi level. Under the influence of the long-range moir\'e potential, the Fermi surface develops multiple band inversions and crossing points. The entanglement among these crossing bands generates a significant spin Berry curvature in momentum space. As shown in Fig.~\ref{NbS}(b), NbSe$_2$ exhibits a SHC as large as $-17\,e/4\pi$ at the Fermi level, which becomes as large as \(-5200\ (\hbar/e)\mathrm{S/cm}\) in three-dimensional units. We note this is almost three time of current record SHE in platinum around \(-2000\ (\hbar/e)\mathrm{S/cm}\)~\cite{nguyen2016spin,zhang2015role,guo2008intrinsic,zhang2021different}.

To directly reveal the origin of enhanced SHC in the moir\'e metal regime with large Fermi surface, we unfold the moir\'e bands into the primitive Brillouin zone of monolayer MoTe$_2$ and NbSe$_2$ using the spectral function defined as~\cite{fold1}
\begin{equation}
A(\mathbf{k}, \epsilon) = \sum W(\mathbf{k})  \delta(\epsilon - \epsilon{\mathbf{k}}),
\end{equation}
where the spectral weight can be expressed as
\begin{equation}
A(\mathbf{k},\epsilon) = \frac{1}{\mathcal{N}} \sum_{\alpha \beta \gamma} V_{\mathbf{k} \alpha } V_{\mathbf{k} \beta} e^{i \mathbf{k} \cdot [\mathbf{r}_\gamma - \mathbf{r}_\beta]} S_{\alpha \gamma}(\mathbf{k} ).
\end{equation}
Here, we approximate the Dirac delta function by a Gaussian function, with the smearing parameter \(\eta\) set to 0.02:
$\delta(x) \approx \frac{1}{\sqrt{\pi} \, \eta} \, e^{-x^2 / \eta^2}.$
In Fig.~\ref{sberry}(a), we present the folded Fermi surface of t-MoTe$_2$ at a twist angle of $5.09^\circ$, with the Fermi energy located 170\,meV above the conduction band minimum. A large Fermi surface emerges near the Brillouin zone boundary, where band folding leads to partial fragmentation of the originally continuous Fermi surface. The fragmentation arises as bands crossing the Fermi level become gapped by the moir\'e potential, breaking the continuity of the Fermi surface and yielding a broad distribution of spin Berry curvature. Correspondingly, as illustrated in Fig.~\ref{sberry}(b), significant spin Berry curvature appears near the Brillouin zone edge, contributing to a spin Hall conductivity of $6\tfrac{e}{4\pi}$. In Fig.~\ref{sberry}(c), we present the folded Fermi surface of t-NbSe$_2$ at a twist angle of $5.09^\circ$, with the Fermi energy located at the charge neutrality point. Due to the relatively large Fermi surface area in this metallic system, more states on the Fermi surface are gapped out by the moir\'e potential, thereby contributing significantly to the spin Berry curvature. Figure~\ref{sberry}(d) shows the spin Berry curvature of t-NbSe$_2$ at $5.09^\circ$, with large curvature appearing at the chemical potential where the Fermi surface is gapped and dense band-inversion networks are created by long-wavelength moir\'e potentials.

In summary, we employ GPU-accelerated large-scale \textit{ab-initio} quantum transport simulations to reveal universal spin Hall effect enhancement mechanisms across moir\'e semiconductors and metals. For t-MoTe$_2$, we demonstrate a dual amplification pathway: (1) In lightly doped regimes, isolated topological flat bands produce quantized SHC reaching $10\,\frac{e}{4\pi}$ at $1.89^\circ$, while (2) heavy doping induces metallic states where moir\'e-driven Fermi surface reconstruction triples the peak SHC to $17\,\frac{e}{4\pi}$ at $3.89^\circ$---surpassing values from topological bands. Remarkably, in intrinsic moir\'e metals ($5.09^\circ$ t-NbS$_2$/t-NbSe$_2$), we discover a colossal SHC of $-17\,\frac{e}{4\pi}$ (-5200 $(\hbar / e)$$S/cm$ in 3D units) at Fermi level, significantly surpassing the current record in Platinum. This enhancement arises from dense band-inversion networks created by long-wavelength moir\'e potentials, which amplify spin Berry curvature across extended Fermi surfaces. Our findings establish moir\'e engineering as a versatile platform for giant spin current generation and general topology-related transport, with metallic systems offering particularly robust signatures for experimental detection.

\section*{Acknowledgments}

We thanks Mark Gates, Natalie Beams, Ahmad Abdelfattah and Jack Dongarra for helpful discussions on large matrix diagonalization. N.M. acknowledges the financial support from the Alexander von Humboldt Foundation. Y.Z. was supported by the start up fund at University of Tennessee.




\bibliography{moire_MoTe}
\clearpage
\pagebreak
\onecolumngrid
\begin{center}
\textbf{\large Supplemental Materials}
\end{center}
\setcounter{figure}{0}
\renewcommand{\thefigure}{S\arabic{figure}}
\setcounter{equation}{0}
\renewcommand{\theequation}{S\arabic{equation}}
\setcounter{table}{0}
\renewcommand{\thetable}{S\arabic{table}}

\section{Transfer learning lattice relaxation}
We adopted a two-step transfer learning approach to model lattice relaxation. In the first stage, a DeepPot-SE descriptor-based model~\cite{zhang2018deep} was trained using a dataset of 6,200 structures, which were generated from 200 random perturbations applied to 28 intermediate transition states. This diverse dataset ensured a comprehensive exploration of the potential energy landscape. In the second stage, by freezing the embedding layer and fine-tuning only the hidden and output layers, the model was specifically optimized for the twisted MoTe$_2$ systems, successfully accommodating twist angles ranging from 5.89$^{\circ}$ to 1.89$^{\circ}$.

\section{OpenMX Calculations}
For the OpenMX calculations, we employed highly efficient basis sets tailored for MoTe$_2$~\cite{openmx_basis,openmx_largescale,openmx_pseudopotential}. The Mo7.0-s3p2d1 basis set, incorporating 14 atomic orbitals, was used for Mo, while the Te7.0-s3p2d2 basis set, with 19 atomic orbitals, was utilized for Te. Using the Perdew-Burke-Ernzerhof (PBE) functional in combination with norm-conserving pseudopotentials, we computed the overlap matrices for the twisted MoTe$_2$ structures with twist angles between 5.09$^{\circ}$ and 1.89$^{\circ}$. This strategy provided accurate results at an exceptionally low computational cost.

\section{Details of DOS calculations}
The DOS is defined by the expression:
\begin{equation}
D(E) = \frac{N_e}{(2\pi)^2} \sum_n \int_{\text{BZ}} \delta(E - \epsilon_{n,\mathbf{k}})\, d^2k,
\end{equation}
where $N_e$ is the band occupancy, and $\delta(E - \epsilon_{n,\mathbf{k}})$ is the delta function centered at energy $E$. In practice, the delta function is approximated using a Gaussian broadening function:
\begin{equation}
\delta(E - E_n) \approx \frac{1}{\sqrt{2\pi\sigma^2}} \exp\left(-\frac{(E - E_n)^2}{2\sigma^2}\right),
\end{equation}
with $\sigma=0.5$ meV.

\section{Edge state calculations}
\begin{figure}[t]
\includegraphics[width=1\columnwidth]{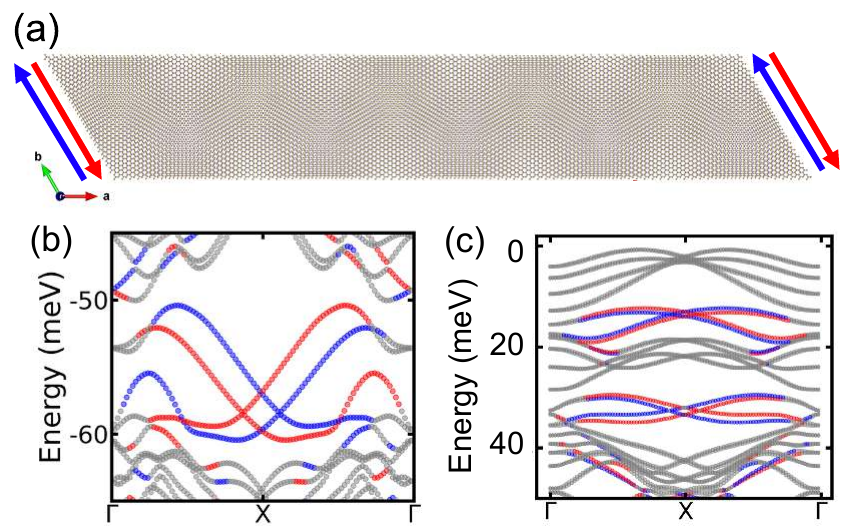}
\caption{
(a) Illustration of the nanoribbon, which is periodic along the b direction and finite along the a direction, spanning a length of 5 unit cells. The red arrows and blue arrows denote for the flow of spin-up and spin-down channels, respectively. The band structure of nanoribbons for twist angle of (b) 5.09$^\circ$ and (c) 3.89$^\circ$, where the edge states of spin-up/-down are in red/blue, and the bulk states are in gray.
}\label{edge}
\end{figure}

The spin Chern number determines whether a system supports topologically protected edge states under open boundary conditions. Furthermore, its magnitude corresponds to the number of helical edge states in the system. Therefore, we construct the Hamiltonian and overlap matrices for a ribbon geometry and perform sparse diagonalization to explore the multiple quantum spin Hall states. The bulk Hamiltonian and overlap matrices include nine distinct hopping types, defined by the distances
$\mathbf{R}_{1-9}$ = [0 0 0], [0 1 0], [0 -1 0], [1 0 0], [-1 0 0], [1 1 0], [-1 -1 0], [1 -1 0], and [-1 1 0].
Among these, \(\mathbf{R}_{5\text{-}9}\) are used for interlayer hoppings, while \(\mathbf{R}_{1\text{-}4}\) are used for intralayer hoppings. We then define the ribbon Hamiltonian and overlap matrix as
\begin{equation}
H/S^{\mathrm{slab}}(\mathbf{k})
=
\begin{pmatrix}
H/S^{11}(\mathbf{k}) & H/S^{12}(\mathbf{k}) & \cdots & H/S^{1n}(\mathbf{k})\\
H/S^{21}(\mathbf{k}) & H/S^{22}(\mathbf{k}) & \cdots & H/S^{2n}(\mathbf{k})\\
\vdots & \vdots & \ddots & \vdots \\
H/S^{n1}(\mathbf{k}) & H/S^{n2}(\mathbf{k}) & \cdots & H/S^{nn}(\mathbf{k})
\end{pmatrix},
\end{equation}
where \(n\) labels the layer index, and \(\bigl(H/S\bigr)^{mn}(\mathbf{k})\) represents the corresponding Hamiltonian or overlap matrices for distances \(\mathbf{R}_{mn} = \mathbf{R}_{m} - \mathbf{R}_{n}\). Since only nine hopping types are included, \(\bigl(H/S\bigr)^{mn}(\mathbf{k})\) vanishes whenever \(\lvert m - n\rvert > 1\). The band structures are obtained via Fourier transform. As shown in Figs.~\ref{edge}, multiple edge states emerge within the bulk gap. In particular, for a twist angle of \(3.89^\circ\), one pair of edge states emerges between the first and second bands, while another pair appears between the second and third bands. In contrast, for a twist angle of \(5.09^\circ\), a global gap exists only between the second and third bands, so only a single pair of edge states is observed within this insulating gap.

\section{Topological classification and Wilson loop calculation}
\begin{figure}[htb]
\includegraphics[width=0.9\columnwidth]{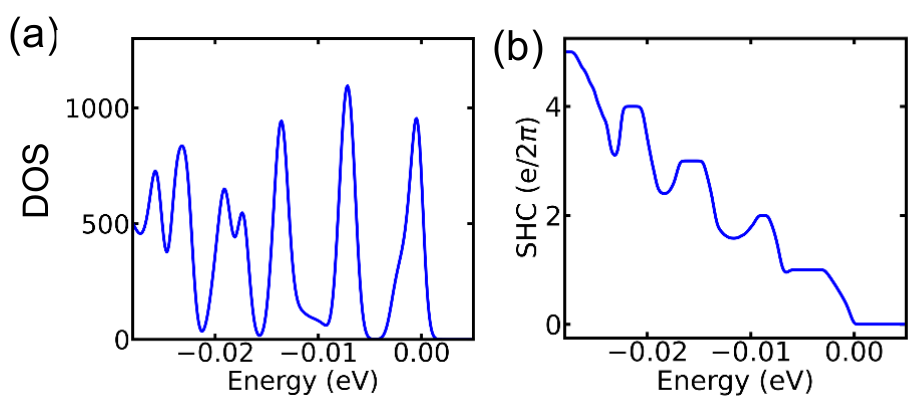}
\caption{ (a) Density of states, and (b) SHC for t-MoTe$_2$ with twist angle 1.89$^{\circ}$.
}\label{189_band}
\end{figure}

\begin{figure}[htb]
\includegraphics[width=0.9\columnwidth]{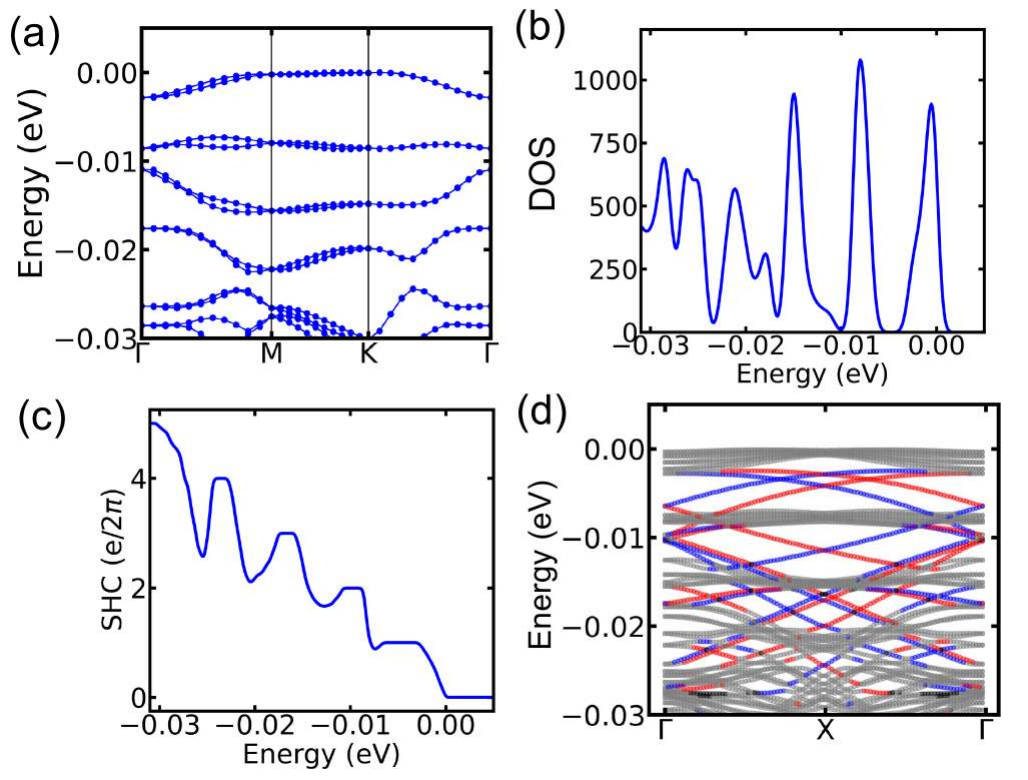}
\caption{ (a) Band structures, (b) density of states, (c) SHC, and (d) edge states for t-MoTe$_2$ with twist angle 2.00$^{\circ}$.
}\label{200_band}
\end{figure}

\begin{figure}[htb]
\includegraphics[width=0.9\columnwidth]{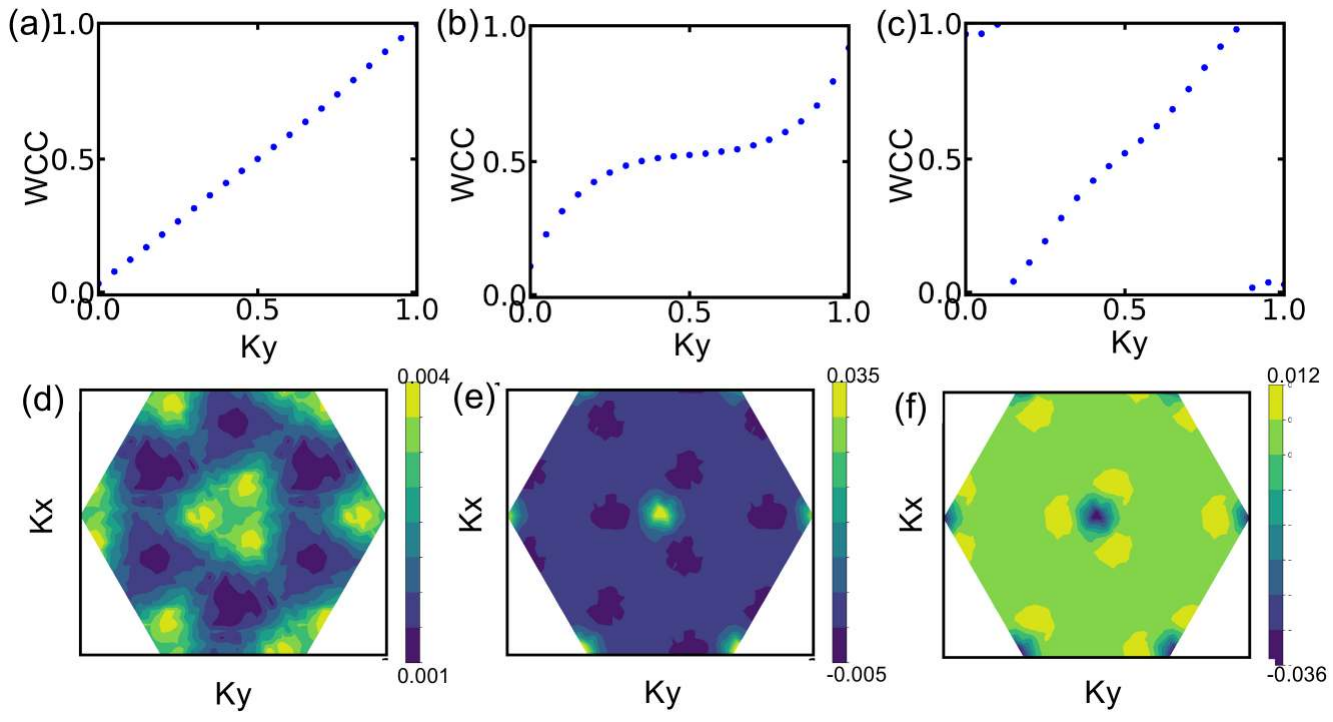}
\caption{The evolution of Wannier charge center and distribution of Berry curvature for the twist angle of 2.00$^{\circ}$, indicating the Chern number of 1, 1, and 1 for the first, second, and third band, respectively. Panels (a) and (d) correspond to the first band, (b) and (e) correspond to the second band, while (c) and (f) correspond to the third band.
}\label{200_topology}
\end{figure}

\begin{figure}[htb]
\includegraphics[width=0.9\columnwidth]{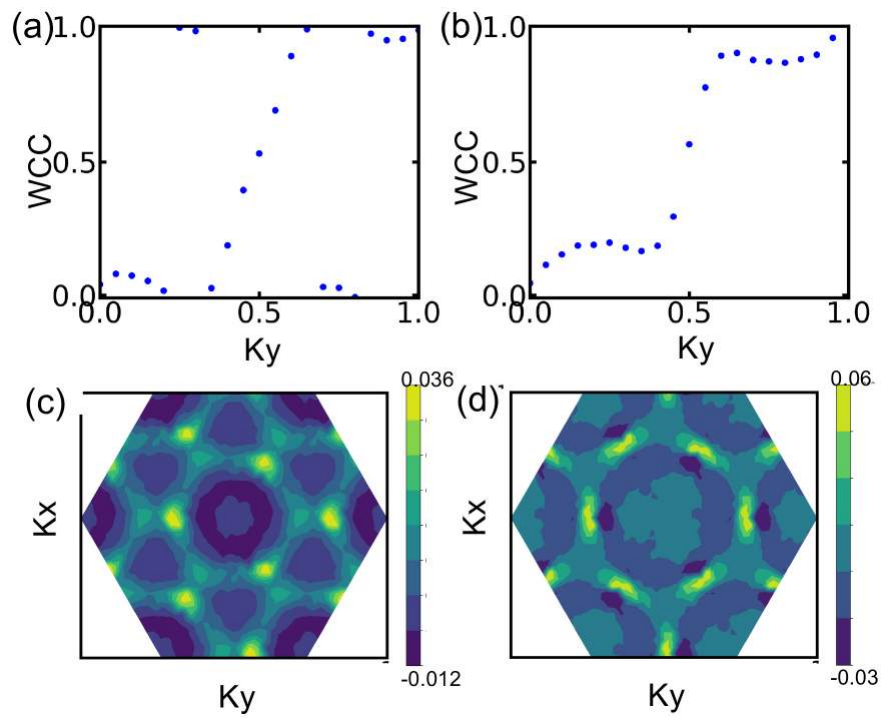}
\caption{The evolution of Wannier charge center and distribution of Berry curvature for the twist angle of 2.00$^{\circ}$, indicating the Chern number of 1 and 1 for the fourth and fifth band, respectively. Panels (a) and (c) correspond to the fourth band, while (b) and (d) correspond to the fifth band.
}\label{200_topology_2}
\end{figure}

\begin{figure}[htb]
\includegraphics[width=0.9\columnwidth]{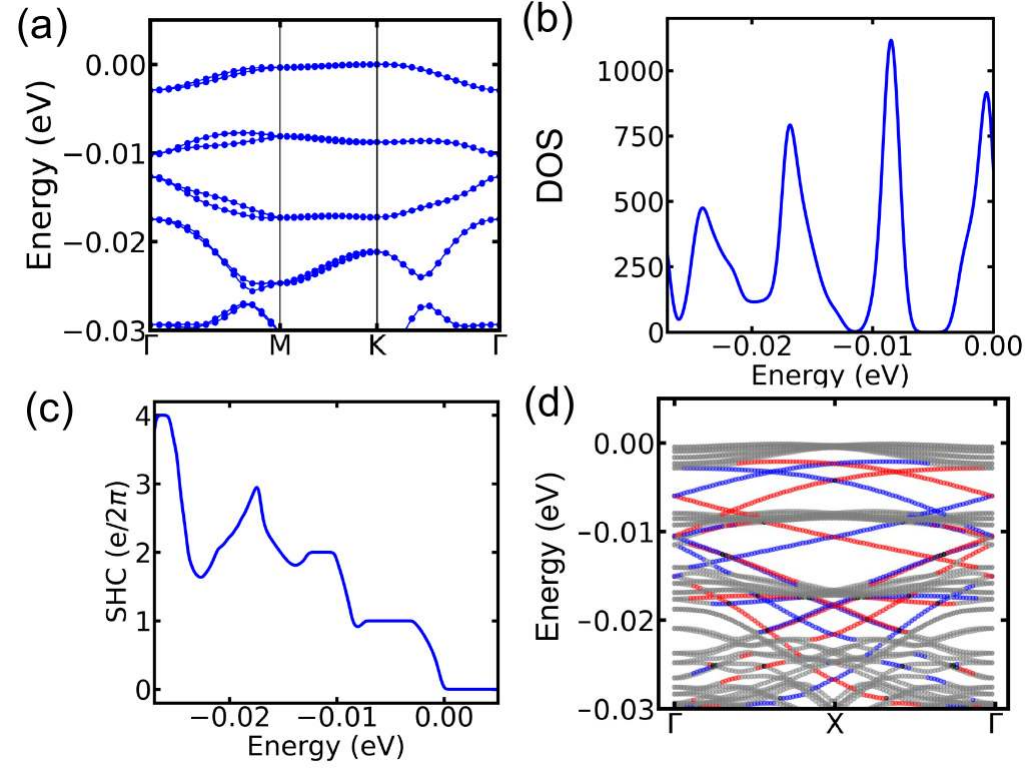}
\caption{ (a) Band structures, (b) density of states, (c) SHC, and (d) edge states for t-MoTe$_2$ with twist angle 2.13$^{\circ}$.
}\label{213_band}
\end{figure}

\begin{figure}[htb]
\includegraphics[width=0.9\columnwidth]{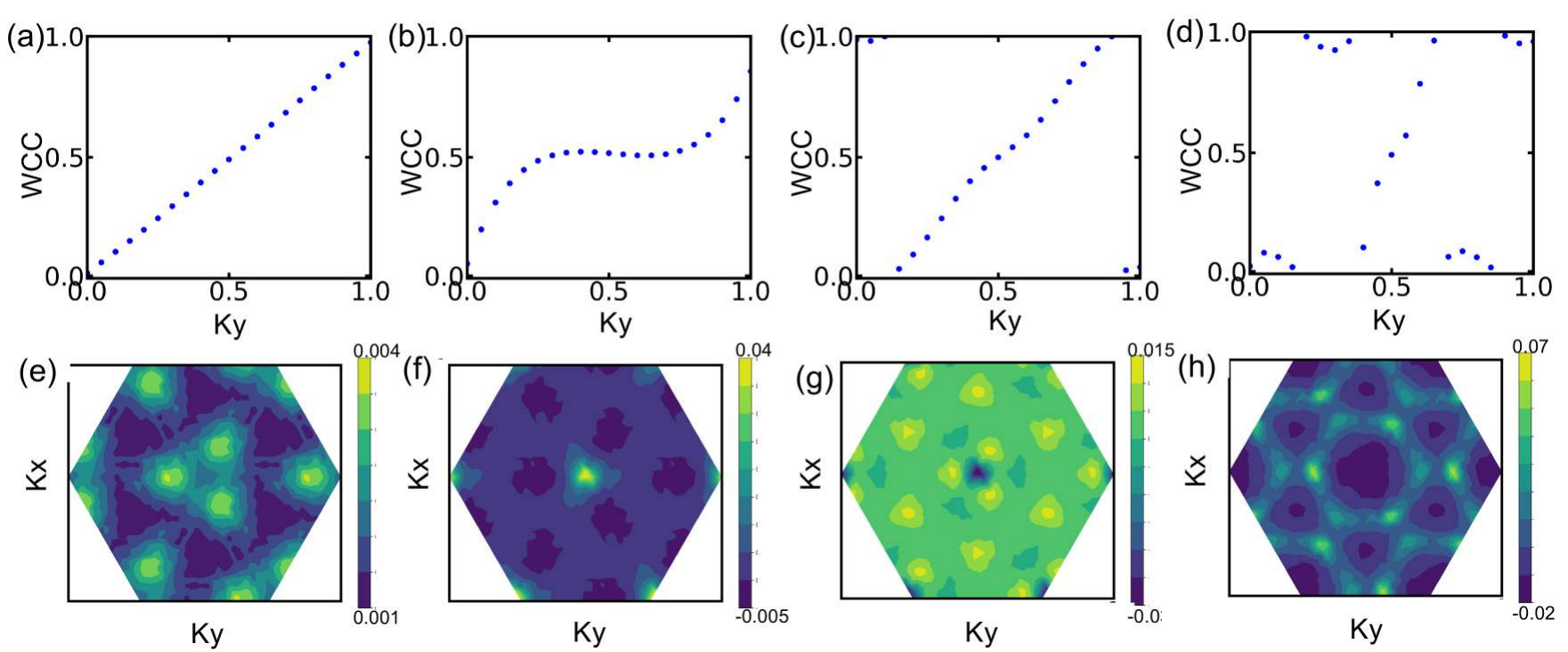}
\caption{The evolution of Wannier charge center and distribution of Berry curvature for the twist angle of 2.13$^{\circ}$, indicating the Chern number of 1, 1, 1, and 1 for the first, second, third, and fourth band, respectively. Panels (a) and (e) correspond to the first band, (b) and (f) correspond to the second band, (c) and (g) correspond to the third band, while (d) and (h) correspond to the fourth band.
}\label{213_topology}
\end{figure}

\begin{figure}[htb]
\includegraphics[width=0.9\columnwidth]{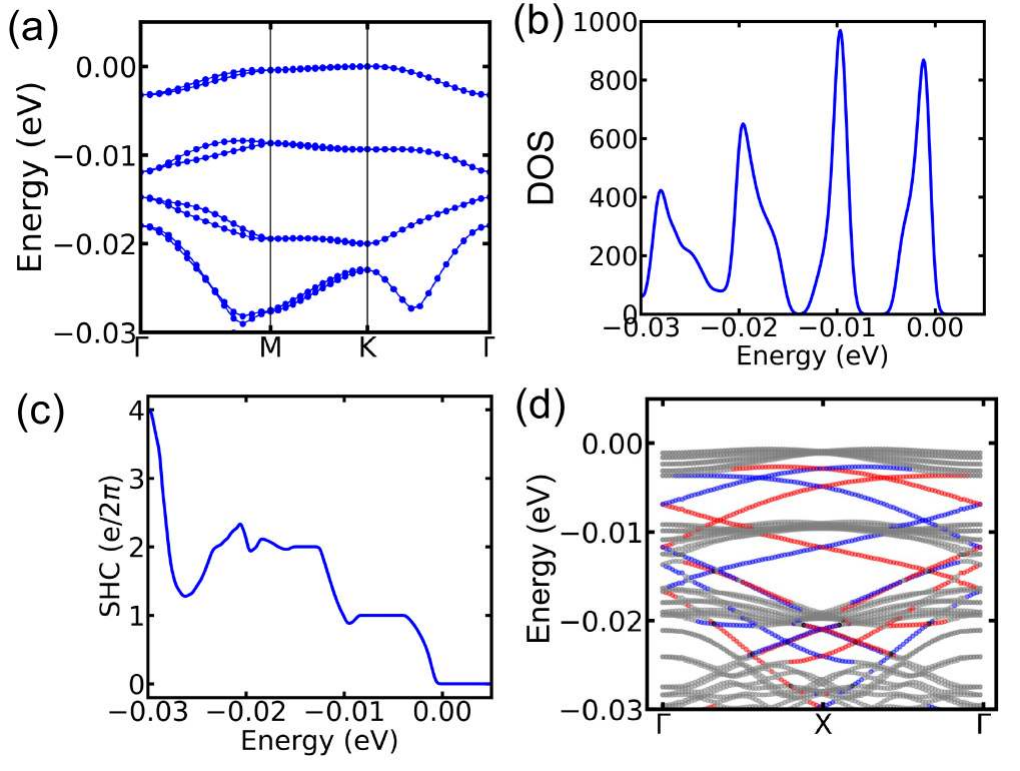}
\caption{ (a) Band structures, (b) density of states, (c) spin Hall conductivity, and (d) edge states for t-MoTe$_2$ with twist angle 2.28$^{\circ}$.
}\label{228_band}
\end{figure}

\begin{figure}[htb]
\includegraphics[width=0.9\columnwidth]{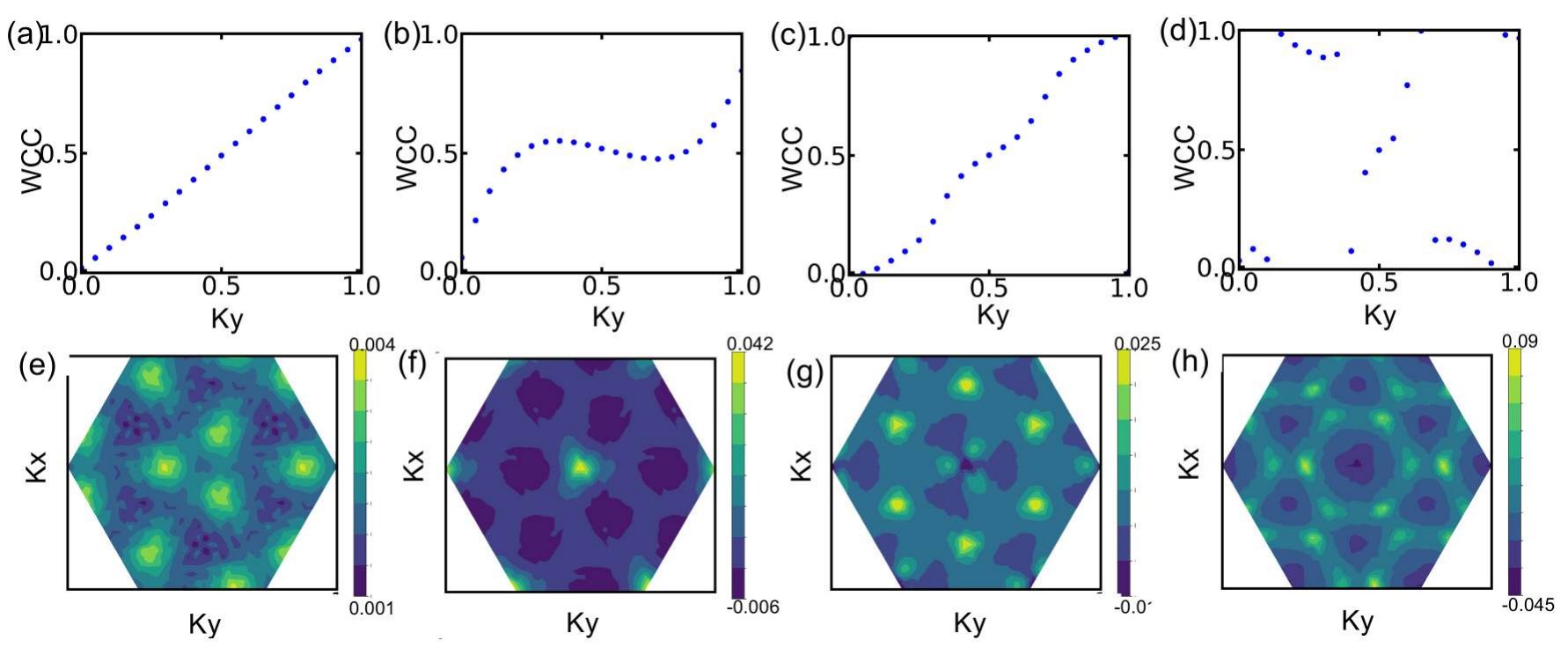}
\caption{The evolution of Wannier charge center and distribution of Berry curvature for the twist angle of 2.28$^{\circ}$, indicating the Chern number of 1, 1, 1, and 1 for the first, second, third, and fourth band, respectively. Panels (a) and (e) correspond to the first band, (b) and (f) correspond to the second band, (c) and (g) correspond to the third band, while (d) and (h) correspond to the fourth band.
}\label{228_topology}
\end{figure}

\begin{figure}[htb]
\includegraphics[width=0.9\columnwidth]{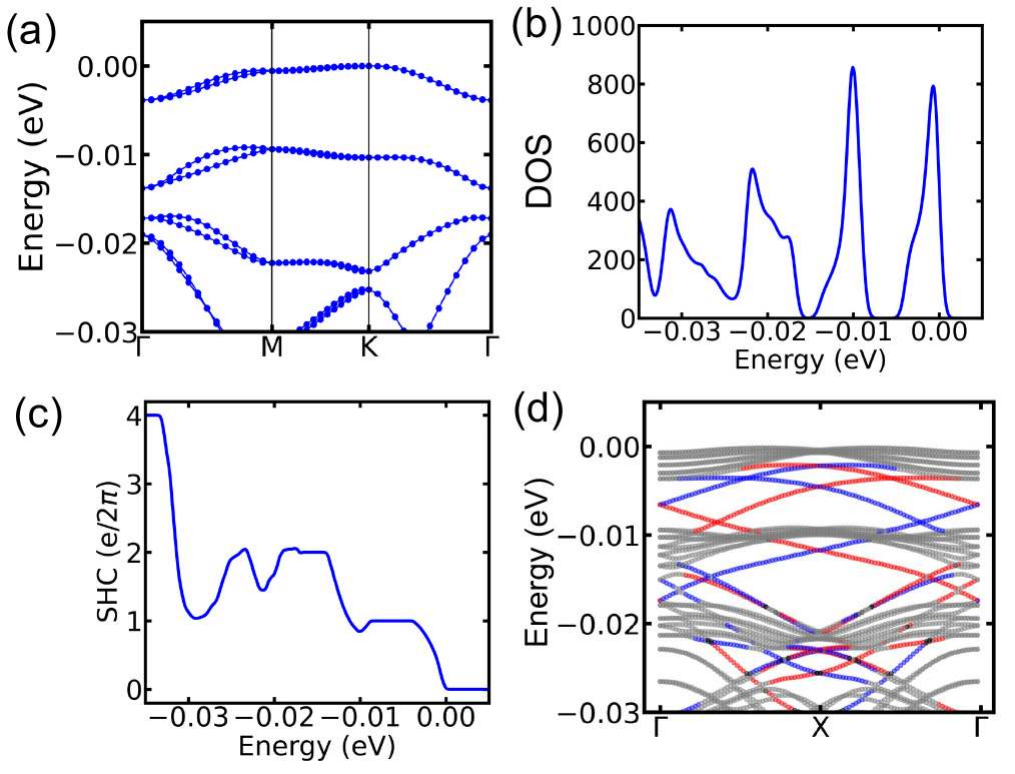}
\caption{ (a) Band structures, (b) density of states, (c) SHC, and (d) edge states for t-MoTe$_2$ with twist angle 2.45$^{\circ}$.
}\label{245_band}
\end{figure}

\begin{figure}[htb]
\includegraphics[width=0.9\columnwidth]{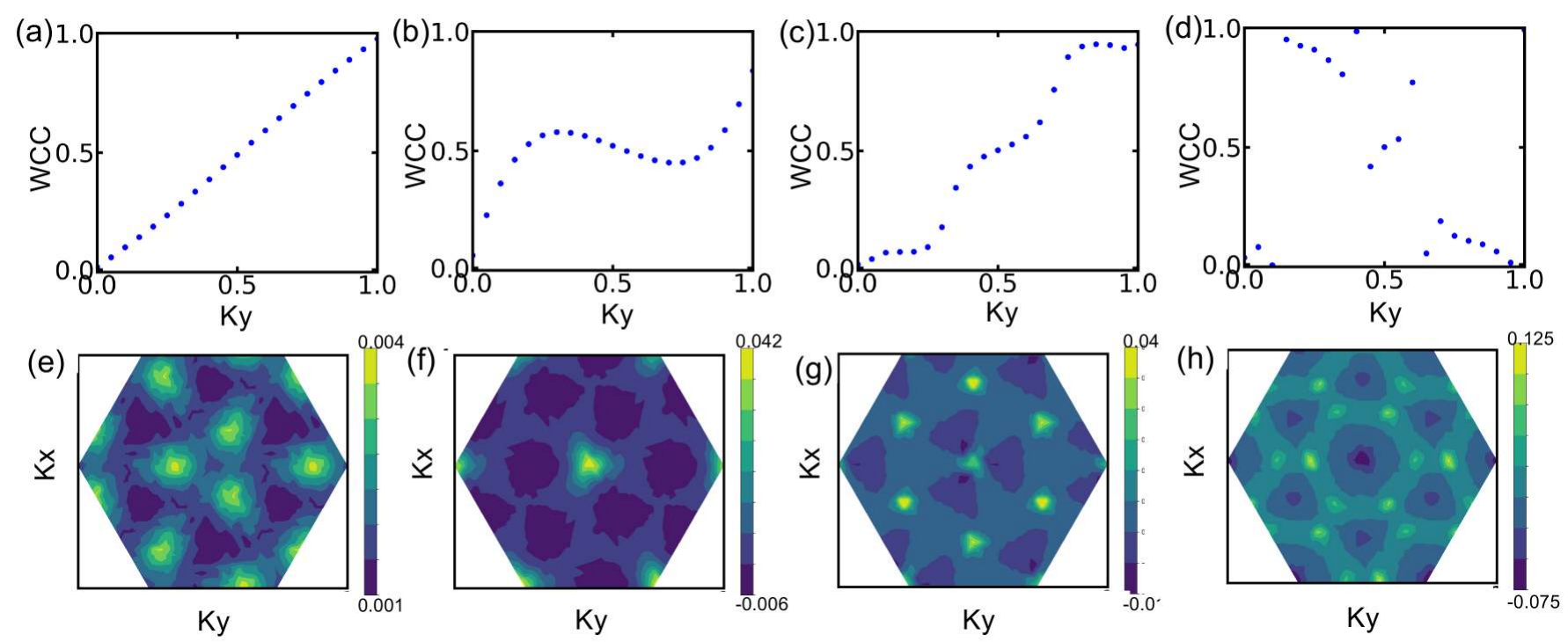}
\caption{The evolution of Wannier charge center and distribution of Berry curvature for the twist angle of 2.45$^{\circ}$, indicating the Chern number of 1, 1, 1, and 1 for the first, second, third, and fourth band, respectively. Panels (a) and (e) correspond to the first band, (b) and (f) correspond to the second band, (c) and (g) correspond to the third band, while (d) and (h) correspond to the fourth band.
}\label{245_topology}
\end{figure}

\begin{figure}[htb]
\includegraphics[width=0.9\columnwidth]{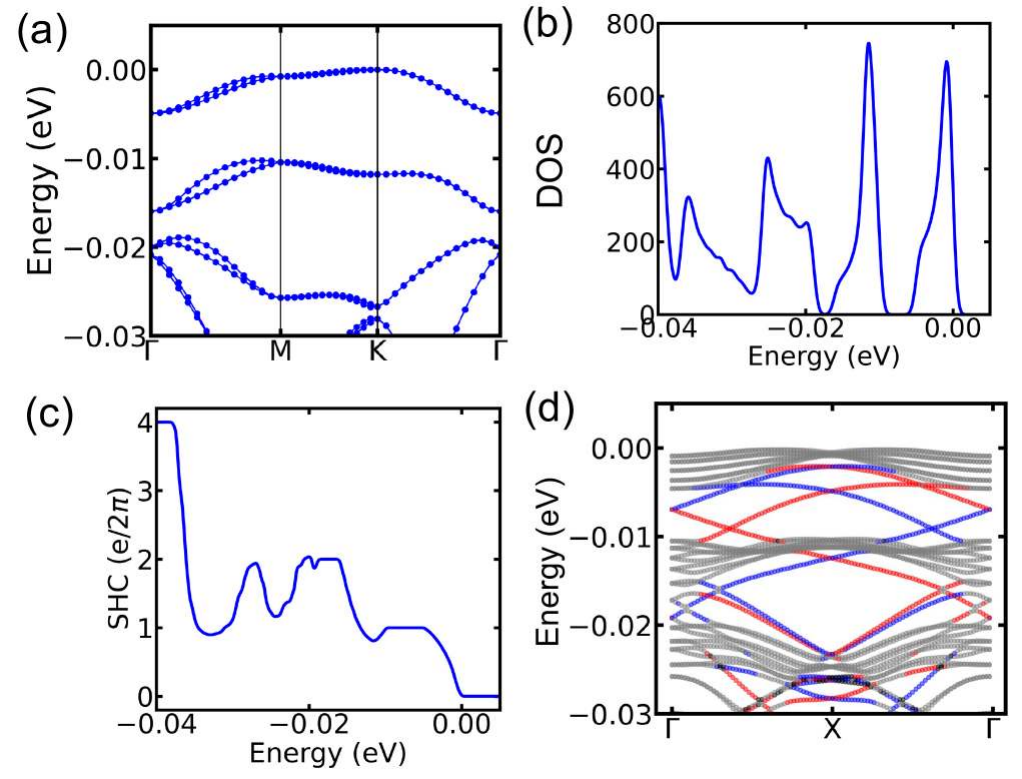}
\caption{(a) Band structures, (b) density of states, (c) SHC, and (d) edge states for t-MoTe$_2$ with twist angle 2.65$^{\circ}$.
}\label{265_band}
\end{figure}

\begin{figure}[htb]
\includegraphics[width=0.9\columnwidth]{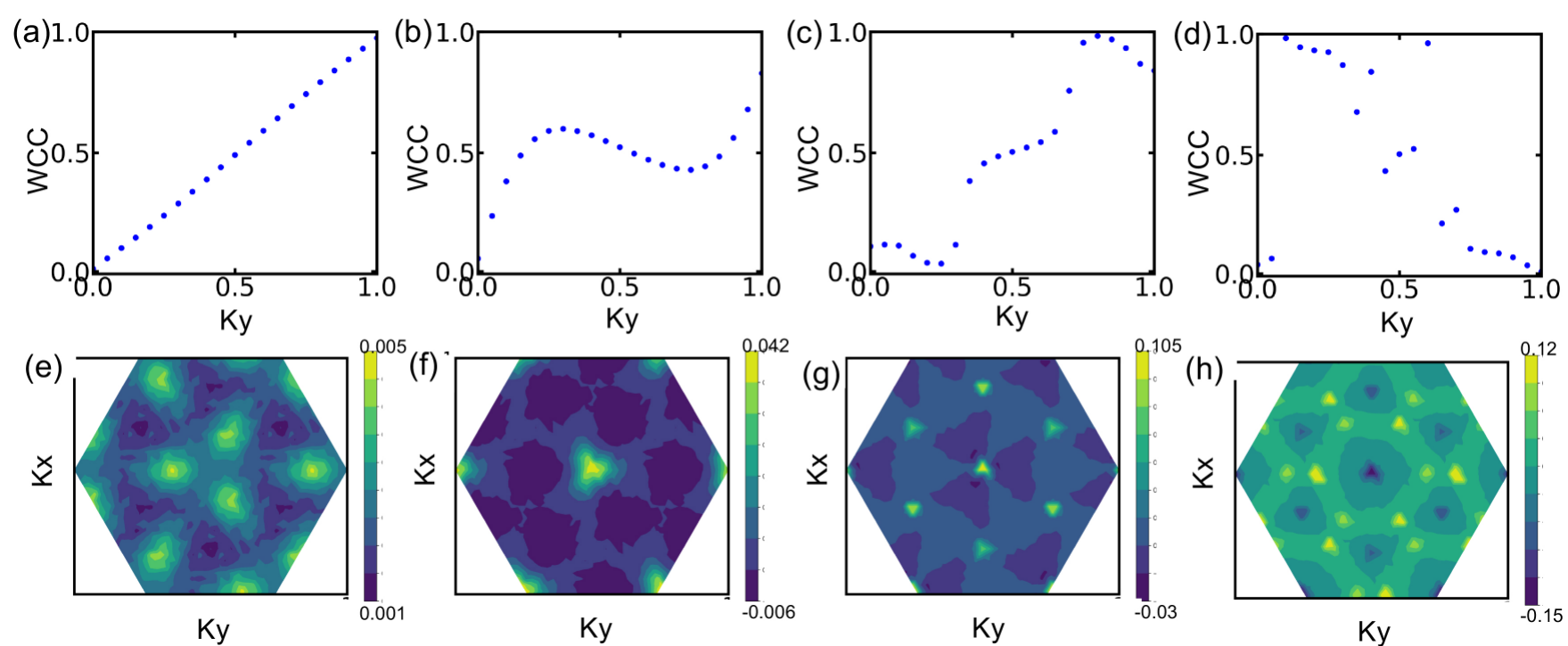}
\caption{The evolution of Wannier charge center and distribution of Berry curvature for the twist angle of 2.65$^{\circ}$, indicating the Chern number of 1, 1, 1, and 1 for the first, second, third, and fourth band, respectively. Panels (a) and (e) correspond to the first band, (b) and (f) correspond to the second band, (c) and (g) correspond to the third band, while (d) and (h) correspond to the fourth band.
}\label{265_topology}
\end{figure}

\begin{figure}[htb]
\includegraphics[width=0.9\columnwidth]{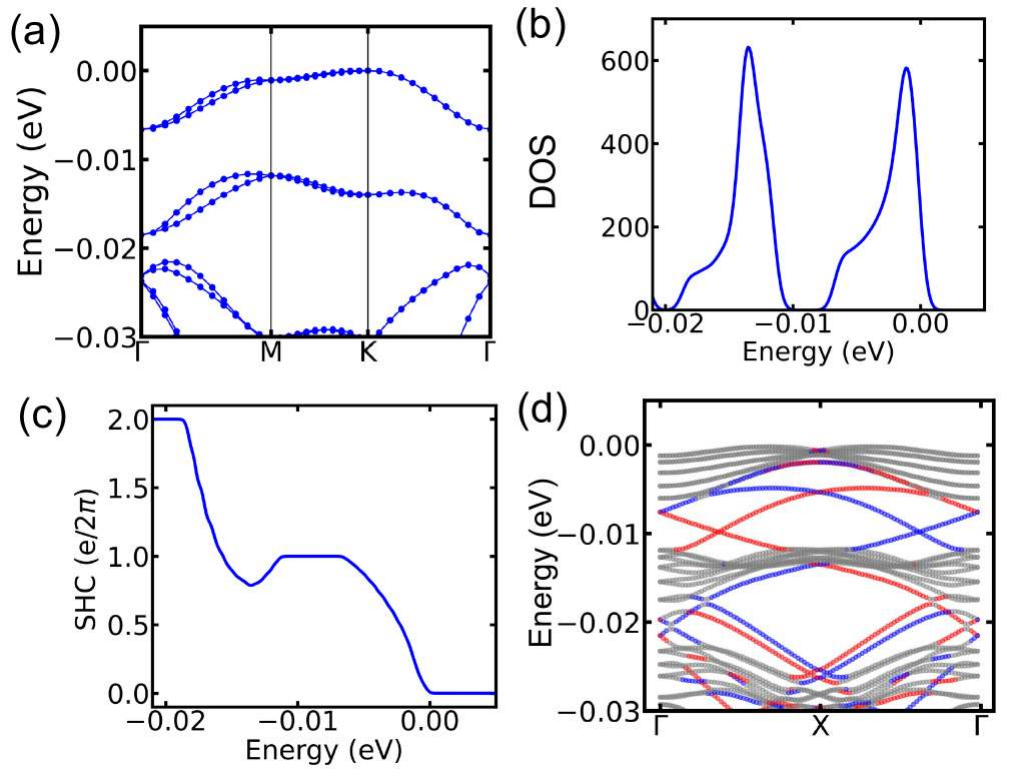}
\caption{ (a) Band structures, (b) density of states, (c) SHC, and (d) edge states for t-MoTe$_2$ with twist angle 2.88$^{\circ}$.
}\label{288_band}
\end{figure}

\begin{figure}[htb]
\includegraphics[width=0.9\columnwidth]{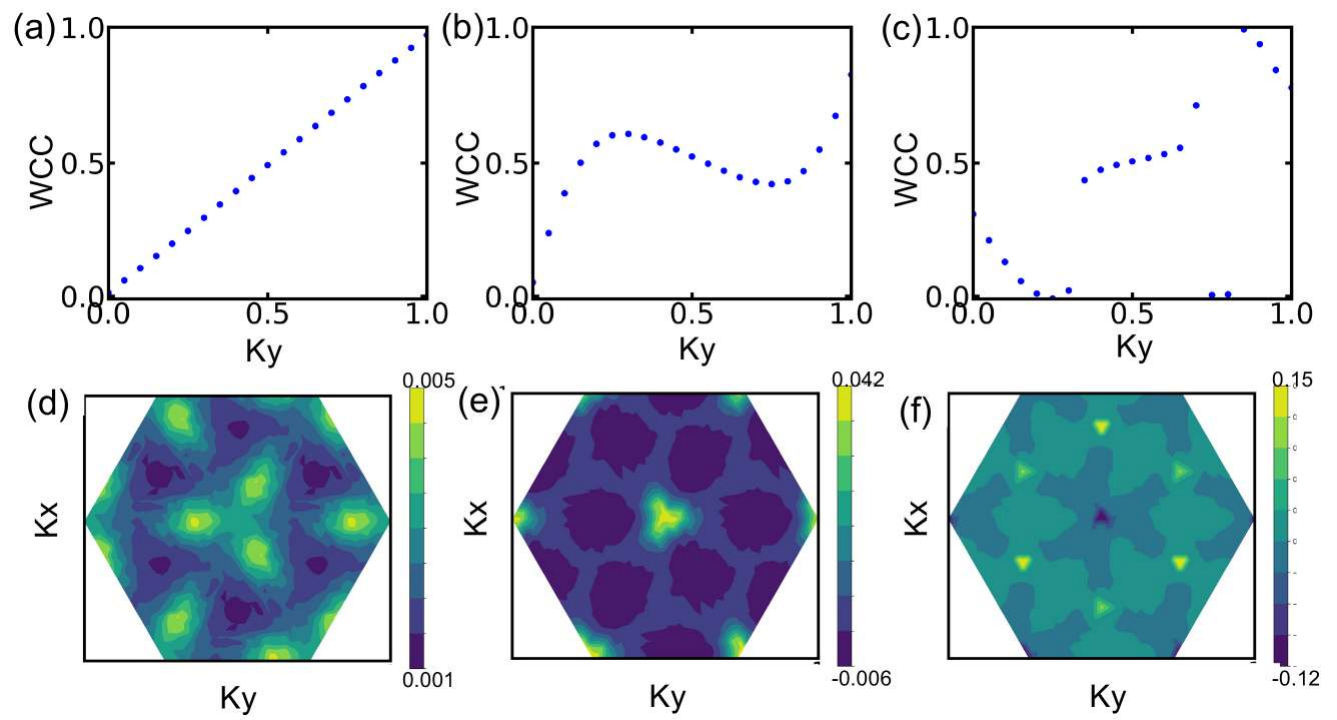}
\caption{The evolution of Wannier charge center and distribution of Berry curvature for the twist angle of 2.88$^{\circ}$, indicating the Chern number of 1, 1, and 0 for the first, second, and third band, respectively. Panels (a) and (d) correspond to the first band, (b) and (e) correspond to the second band, while (c) and (f) correspond to the third band.
}\label{288_topology}
\end{figure}

\begin{figure}[htb]
\includegraphics[width=0.9\columnwidth]{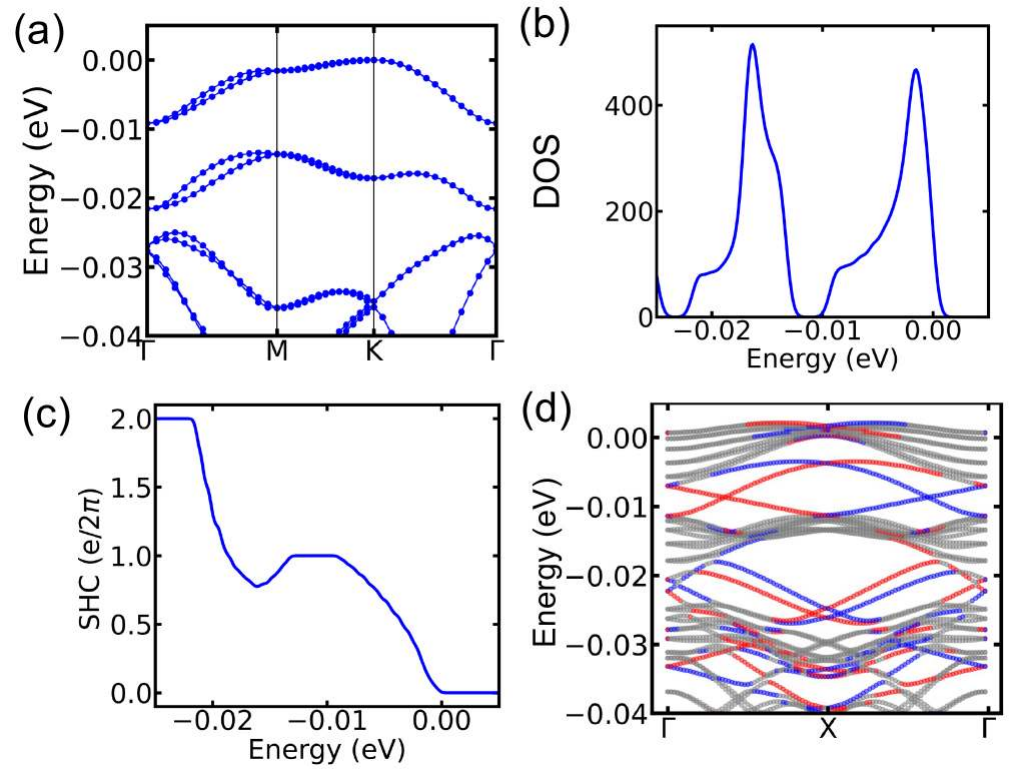}
\caption{ (a) Band structures, (b) density of states, (c) SHC, and (d) edge states for t-MoTe$_2$ with twist angle 3.15$^{\circ}$.
}\label{315_band}
\end{figure}

\begin{figure}[htb]
\includegraphics[width=0.9\columnwidth]{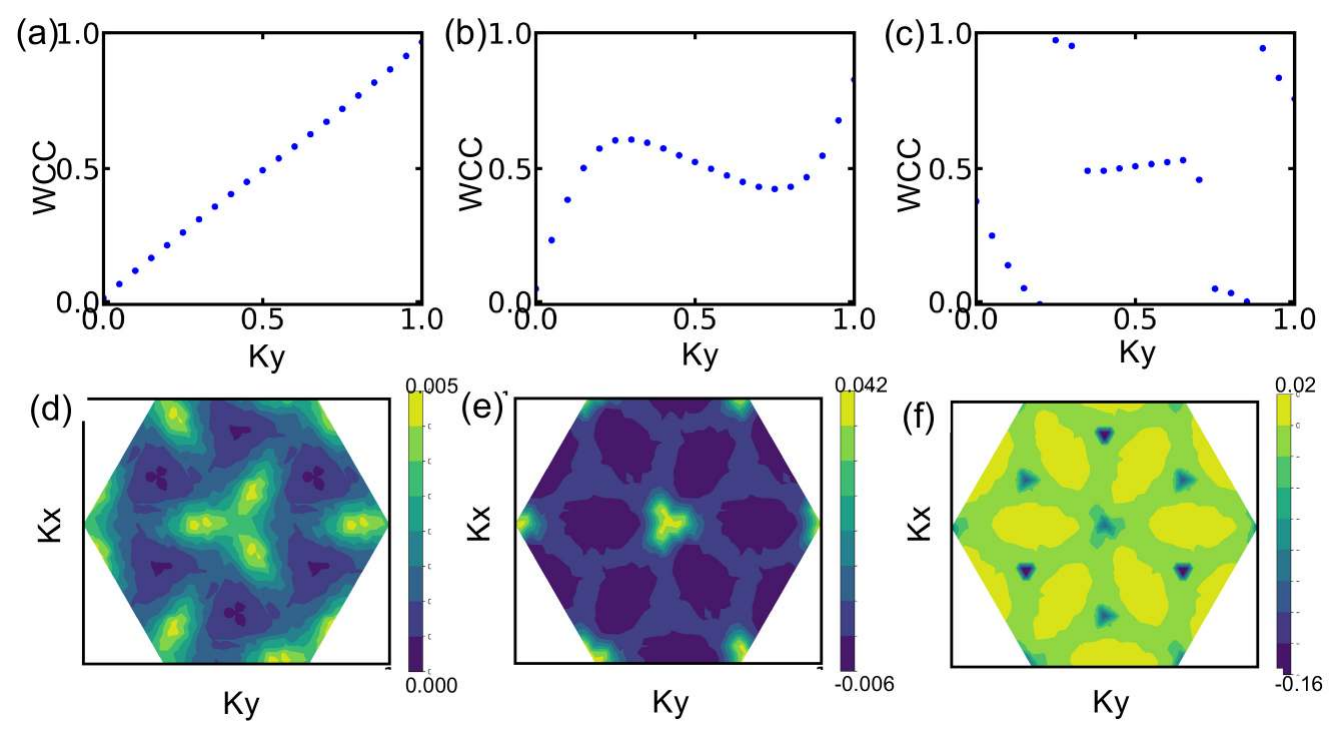}
\caption{The evolution of Wannier charge center and distribution of Berry curvature for the twist angle of 3.15$^{\circ}$, indicating the Chern number of 1, 1, and -2 for the first, second, and third band, respectively. Panels (a) and (d) correspond to the first band, (b) and (e) correspond to the second band, while (c) and (f) correspond to the third band.
}\label{315_topology}
\end{figure}

\begin{figure}[htb]
\includegraphics[width=0.9\columnwidth]{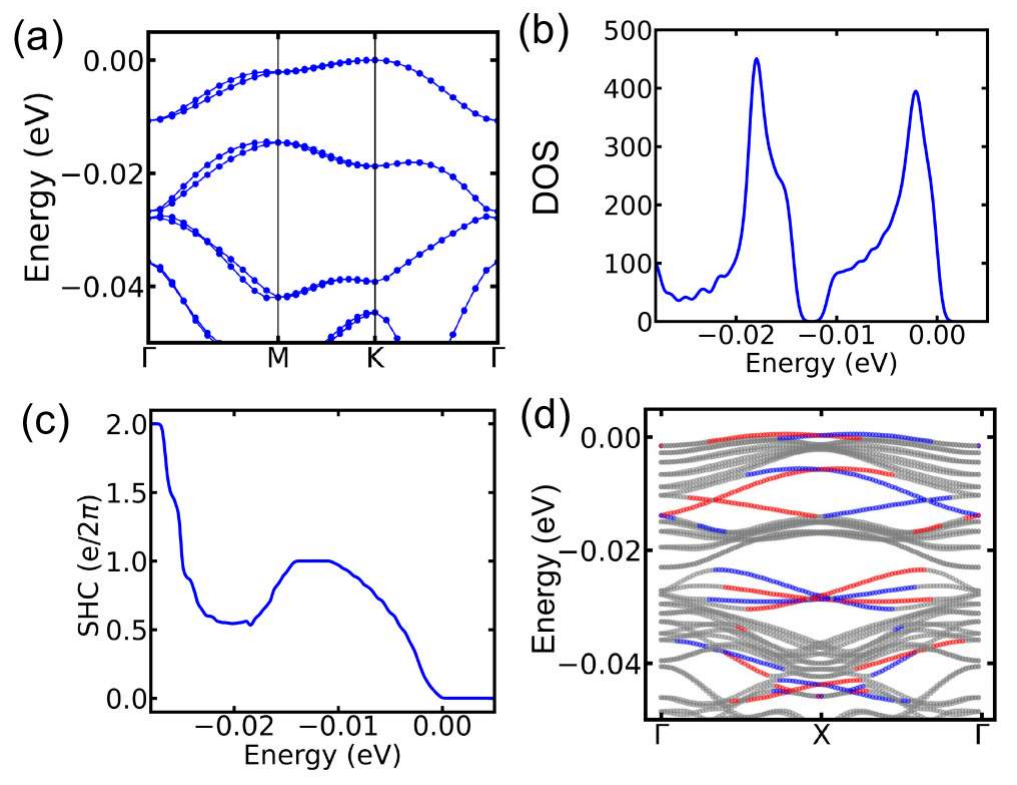}
\caption{ (a) Band structures, (b) density of states, (c) SHC, and (d) edge states for t-MoTe$_2$ with twist angle 3.48$^{\circ}$.
}\label{348_band}
\end{figure}

\begin{figure}[htb]
\includegraphics[width=0.9\columnwidth]{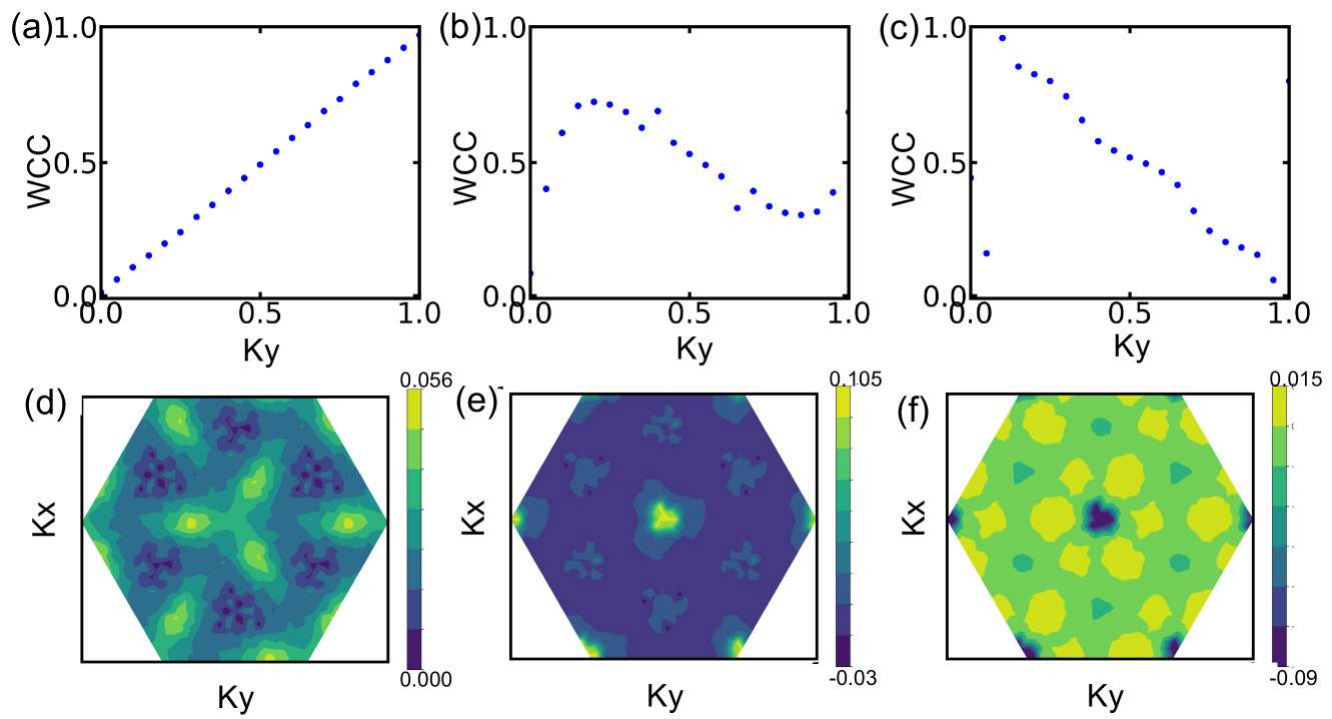}
\caption{The evolution of Wannier charge center and distribution of Berry curvature for the twist angle of 3.48$^{\circ}$, indicating the Chern number of 1, 1, and -2 for the first, second, and third band, respectively. Panels (a) and (d) correspond to the first band, (b) and (e) correspond to the second band, while (c) and (f) correspond to the third band.
}\label{348_topology}
\end{figure}

\begin{figure}[htb]
\includegraphics[width=0.9\columnwidth]{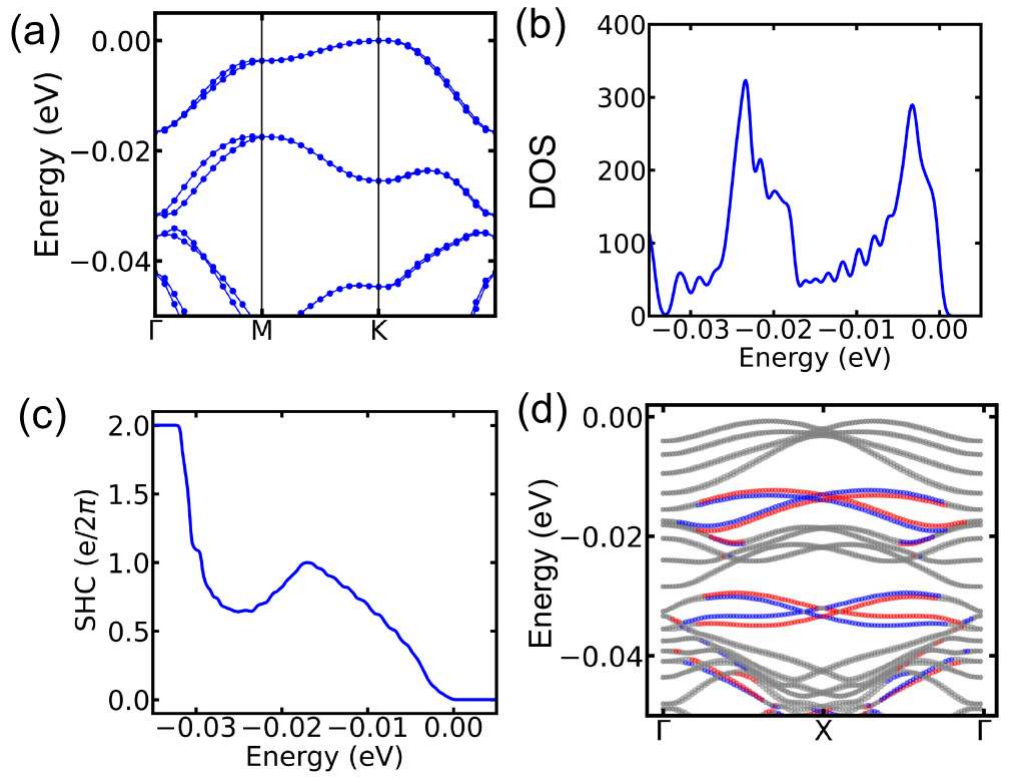}
\caption{ (a) Band structures, (b) density of states, (c) SHC, and (d) edge states for t-MoTe$_2$ with twist angle 3.89$^{\circ}$.
}\label{389_band}
\end{figure}

\begin{figure}[htb]
\includegraphics[width=0.9\columnwidth]{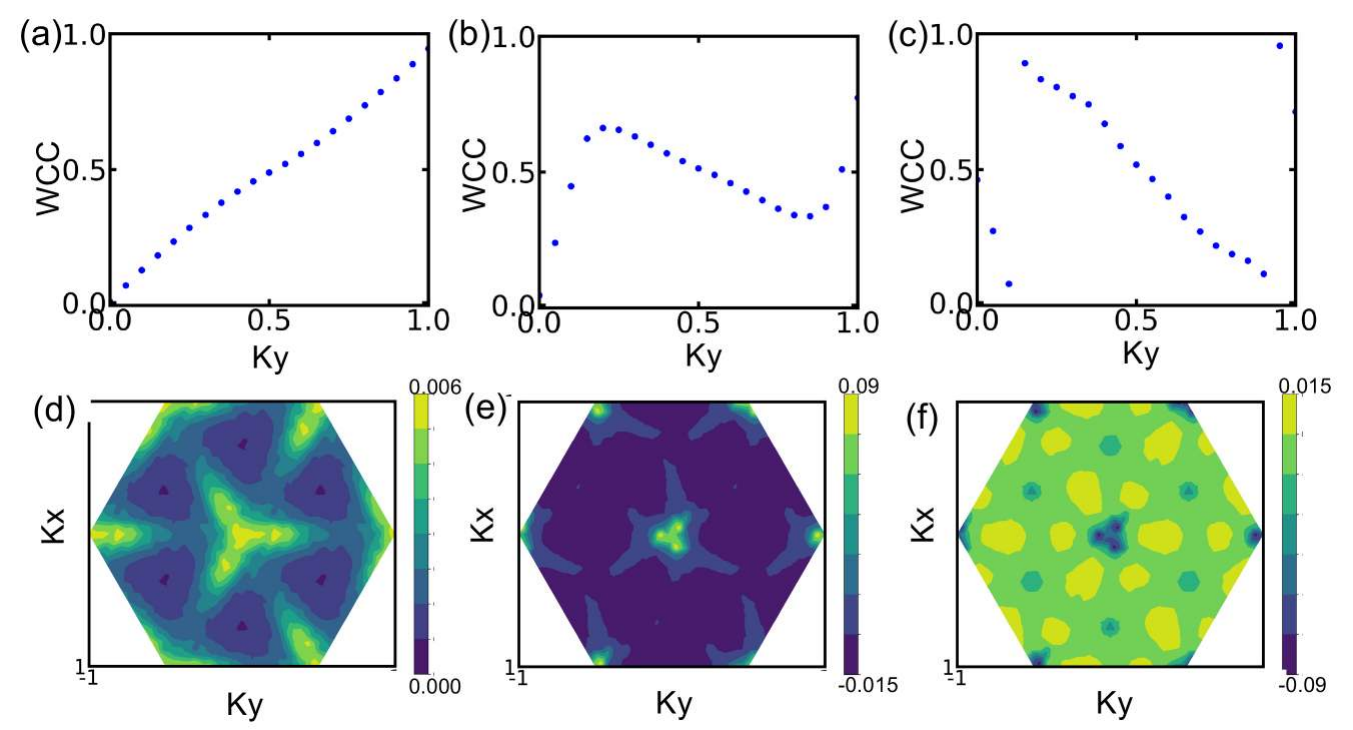}
\caption{The evolution of Wannier charge center and distribution of Berry curvature for the twist angle of 3.89$^{\circ}$, indicating the Chern number of 1, 1, and -2 for the first, second, and third band, respectively. Panels (a) and (d) correspond to the first band, (b) and (e) correspond to the second band, while (c) and (f) correspond to the third band.
}\label{389_topology}
\end{figure}

\begin{figure}[htb]
\includegraphics[width=0.9\columnwidth]{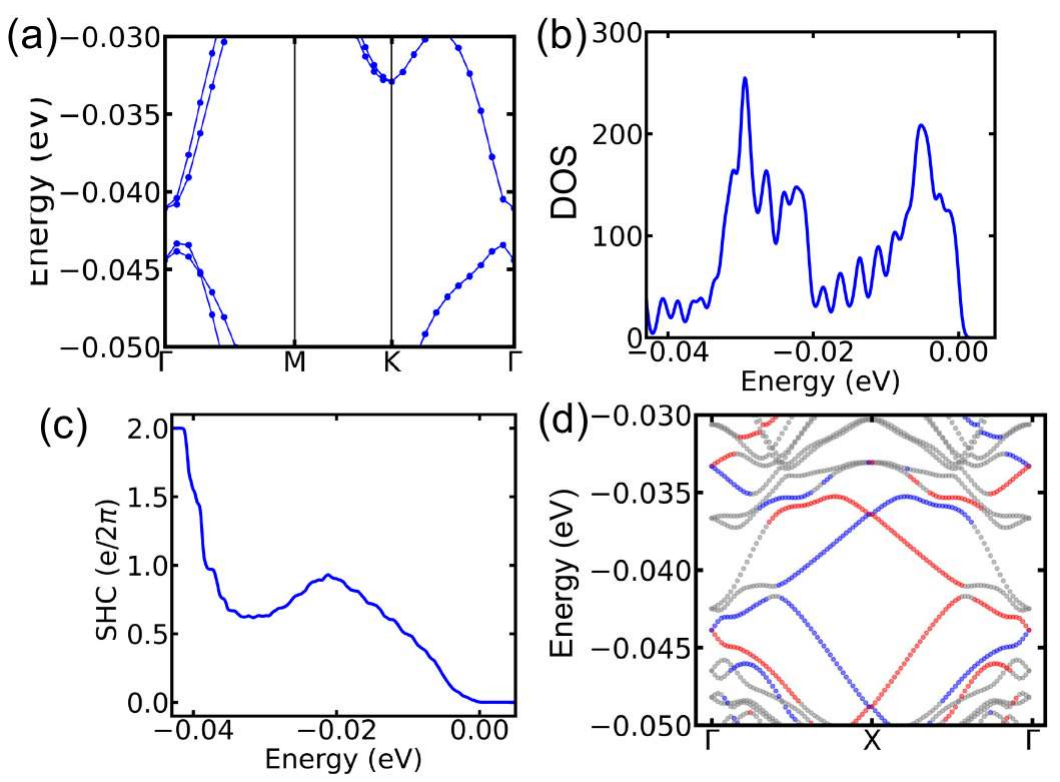}
\caption{ (a) Band structures, (b) density of states, (c) SHC, and (d) edge states for t-MoTe$_2$ with twist angle 4.41$^{\circ}$.
}\label{441_band}
\end{figure}

\begin{figure}[htb]
\includegraphics[width=0.9\columnwidth]{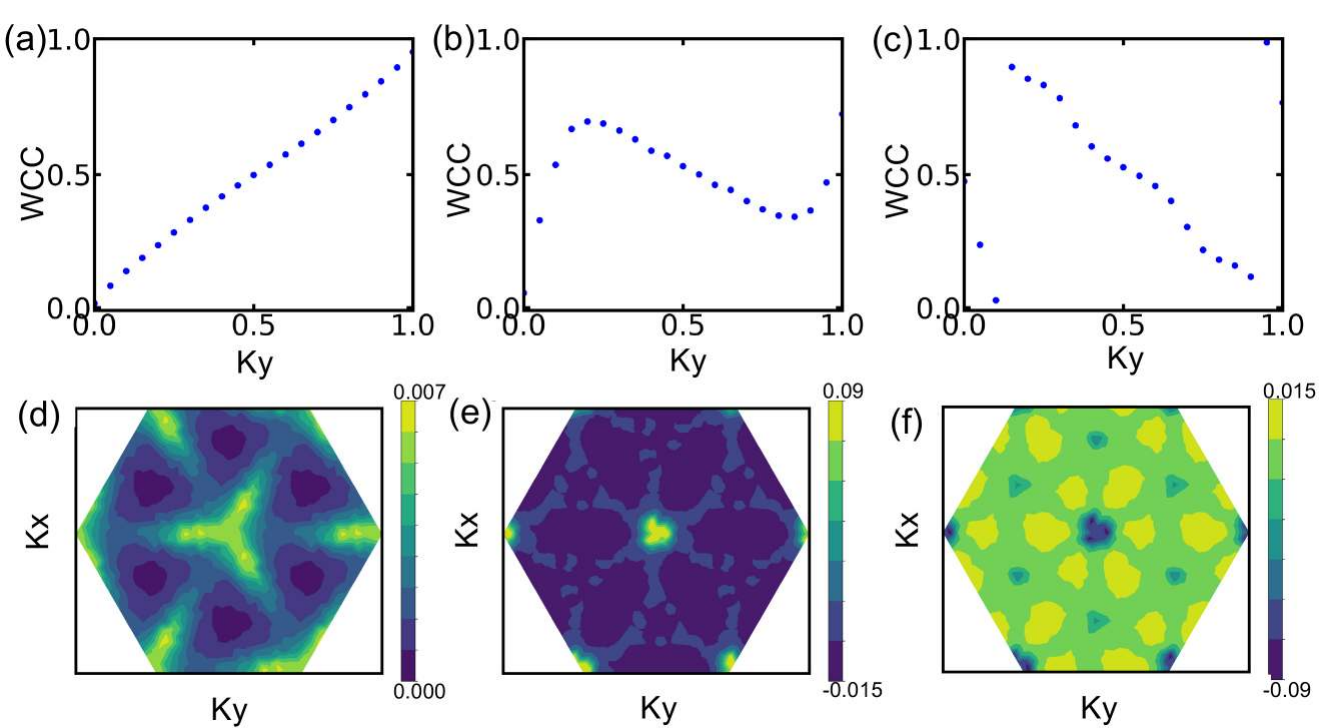}
\caption{The evolution of Wannier charge center and distribution of Berry curvature for the twist angle of 4.4$^{\circ}$, indicating the Chern number of 1, 1, and -2 for the first, second, and third band, respectively. Panels (a) and (d) correspond to the first band, (b) and (e) correspond to the second band, while (c) and (f) correspond to the third band.
}\label{441_topology}
\end{figure}

A convenient approach to analyze topological properties is to perform a Wilson loop calculation, which can achieve convergence even with a relatively small $k$-mesh (e.g., $20\times 20$ in our case). Following Refs.~\cite{wilson1,wilson2}, we begin by computing the expectation value of the position operator through
\begin{equation}
X \;=\; \sum_{n,m} \sum_{\alpha,\,k_x}
V_{n\alpha}^*(k_x)\,V_{m\alpha}(k_x + \delta k_x)\,
\bigl\langle \Psi_{n,k_x} \,\bigl|\,
\Psi_{m,\,k_x + \delta k_x} \bigr\rangle,
\end{equation}
where $\delta k_x$ denotes the discrete spacing along the $k_x$ axis. For convenience, we introduce the overlap matrix,
\begin{equation}
F_{m,n}(\delta \boldsymbol{k}) \;=\; \sum_{\alpha}\, V_{m\alpha}^*(\boldsymbol{k})\,V_{n\alpha}\bigl(\boldsymbol{k} + \delta \boldsymbol{k}\bigr),
\end{equation}
which simplifies the eigenvalue problem. By employing the Wilson loop approach, we construct the matrix product
\begin{equation}
D\bigl(k_y\bigr)
\;=\;
F_{0,1}\,F_{1,2}\,F_{2,3}\,\cdots\,F_{N_{x}-2,\,N_{x}-1}\,F_{N_{x}-1,\,0},
\end{equation}
The eigenvalues $d$ of $D\bigl(k_y\bigr)$ are then used to extract the Wannier charge centers (WCC) via
\begin{equation}
\nu\bigl(k_y\bigr) \;=\; \mathrm{Im}\bigl[\log\,d\bigl(k_y\bigr)\bigr].
\end{equation}
Therefore, the Chern number can be determined by counting the Wannier charge center (WCC) crossings along an arbitrary line as $k_y$ varies from 0 to $\pi$.
Our results for various twist angles are presented in Figs.~\ref{189_band}–\ref{441_topology}. In particular, for twist angles between $5.09^\circ$ and $3.15^\circ$, the Chern number remains $(1,1,-2)$, while at $2.88^\circ$ it becomes $(1,1,0)$. For twist angles between $2.65^\circ$ and $2.13^\circ$, four flat Chern bands with a Chern number of 1 are observed, and at $2.00^\circ$ and $1.89^\circ$, five flat Chern bands with a Chern number of 1 are present.

The topology of two-dimensional time-reversal-invariant insulators is effectively characterized by the $Z_2$ topological invariant~\cite{kane2005quantum,kane2005z}, which distinguishes a topological insulator ($Z_2=1$) from a normal insulator ($Z_2=0$). In addition to the $Z_2$ invariant, the topology can also be captured by the spin Chern number, particularly when the spin-mixing term in the material is negligible~\cite{sheng2005nondissipative}. Unlike the $Z_2$ classification, quantum spin Hall states follow a $Z$ classification~\cite{fukui2007topological,fu2006time}, meaning that the spin Chern number can take any integer value, with its magnitude corresponding to the number of helical edge state pairs. The relationship between the $Z_2$ invariant and the spin Chern number is given by
\begin{equation}
Z_2 = \mathrm{mod}(C_s, 2).
\end{equation}
As a result, small-angle twisted MoTe$_2$, when doped with an odd number of holes, falls into the $Z_2$ nontrivial topological phase, whereas materials doped with an even number of holes belong to the $Z_2$ trivial phase.

\section{Twisted bilayer MoTe$_2$ under hetero-strain}
In twisted bilayer MoTe$_2$, hetero-strain refers to the situation where the two layers experience different strain conditions. This differential strain leads to an anisotropic modification of the moir\'e superlattice, which in turn affects the electronic properties of the system. For simplicity, we refer to the rotated layer as layer 1 and the strained layer as layer 2.
The reciprocal lattice vectors of monolayer can be given by
\begin{equation}
\mathbf{k}_1 = \begin{pmatrix} k \\ 0 \end{pmatrix}, \quad
\mathbf{k}_2 = \begin{pmatrix} k \cos 60^\circ \\ k \sin 60^\circ \end{pmatrix}
\end{equation}
where the magnitude $k$ is defined as
\begin{equation}
k = \frac{4\pi}{\sqrt{3}\,a_0},
\end{equation}
with $a_0 = 3.52\,\text{\AA}$ being the monolayer lattice constant of MoTe$_2$. When layer 1 is twisted by an angle $\theta_T$, the corresponding reciprocal lattice vectors become
\begin{equation}
\mathbf{k}_i' = R(\theta_T)\,\mathbf{k}_i, \quad i = 1,2,
\end{equation}
where the rotation matrix is
\begin{equation}
R(\theta_T) = \begin{pmatrix}
\cos \theta_T & -\sin \theta_T \\
\sin \theta_T & \cos \theta_T
\end{pmatrix}.
\end{equation}
The resulting moir\'e wave vectors are defined as the differences between the reciprocal lattice vectors of the two layers:
\begin{equation}
\mathbf{K}_i = \mathbf{k}_i' - \mathbf{k}_i.
\end{equation}
This leads to the moir\'e wavelength
\begin{equation}
\lambda_M = \frac{a_0}{2\sin\left(\frac{\theta_T}{2}\right)}.
\end{equation}

Next, we apply a strain transformation to layer 2. Here, we only consider uniaxial strain along the $x$-axis. In general, if one wishes to apply uniaxial strain along a direction that makes an angle $\theta_s$ with the $x$-axis, the transformation is given by
\begin{equation}
E'(\epsilon,\theta_s) = R(-\theta_s)E(\epsilon)R(\theta_s).
\end{equation}
The strain matrix and the rotation matrix are
\begin{equation}
\mathbf{E}(\epsilon) =
\begin{pmatrix}
\frac{1}{1+\epsilon} & 0 \\
0 & \frac{1}{1-\delta \epsilon}
\end{pmatrix}, \quad
\mathbf{R}(\theta_s) =
\begin{pmatrix}
\cos\theta_s & -\sin\theta_s \\
\sin\theta_s & \cos\theta_s
\end{pmatrix},
\end{equation}
where $\epsilon$ represents the strain magnitude and $\delta$ is the Poisson ratio of MoTe$_2$. Note that the factor $\frac{1}{1-\delta \epsilon}$ accounts for the Poisson effect.
Then, the modified moir\'e wave vectors are given by
\begin{equation}
\mathbf{K}_i^s = \mathbf{k}_i' - \mathbf{k}_i^s = R(\theta_T)\,\mathbf{k}_i - E'(\epsilon, \theta_s)\,\mathbf{k}_i.
\end{equation}

In our case, since the strain is applied along the $x$-axis, we set $\theta_s=0$.
Moreover, the superlattice vectors for layer 1 are given by the linear combination of the monolayer lattice vectors,
\begin{equation}
\bm{L}_1 = m\,\bm{a}_1+n\,\bm{a}_2, \quad
\bm{L}_2 = p\,\,\bm{a}_1+q\,\,\bm{a}_2,
\end{equation}
and for layer 2 by
\begin{equation}
\bm{L}_1' = m'\,\bm{a}_1'+n'\,\bm{a}_2', \quad
\bm{L}_2' = p'\,\bm{a}_1'+q'\,\bm{a}_2'.
\end{equation}
where $a_1'$ and $a_2'$ are strained lattice vectors.
Due to the strain, the superlattice vectors cannot perfectly match the modified moir\'e wave vectors, resulting in an inherent error. To minimize this error, we employed a gradient descent method by setting the twist angle $\theta_T$ in the range of $2^\circ\sim3^\circ$, and eventually determined the optimal parameters $(m,n,p,q,m',n',p',q')$.

Specifically, we obtained the following configurations:
\begin{itemize}
    \item For $\theta_T \approx 3.676^\circ$, with a uniaxial strain of $0.006$ and an angle between $\bm{a}_1$ and $\bm{a}_2$ of $114^\circ$, the optimal parameters are
    \[
    (m,n,p,q,m',n',p',q') = (9,17,-18,-7,\,8,17,-18,-8).
    \]
    \item For $\theta_T \approx 3.305^\circ$, with a uniaxial strain of $0.005$ and an angle between $\bm{a}_1$ and $\bm{a}_2$ of $115^\circ$, the optimal parameters are
    \[
    (m,n,p,q,m',n',p',q') = (10,19,-20,-8,\,9,19,-20,-9).
    \]
    \item For $\theta_T \approx 2.05^\circ$, with a uniaxial strain of $0.01$ and an angle between $\bm{a}_1$ and $\bm{a}_2$ of $103^\circ$, the optimal parameters are
    \[
    (m,n,p,q,m',n',p',q') = (14,28,-33,-8,\,13,28,-33,-9).
    \]
\end{itemize}

\begin{figure}[htb]
\includegraphics[width=0.9\columnwidth]{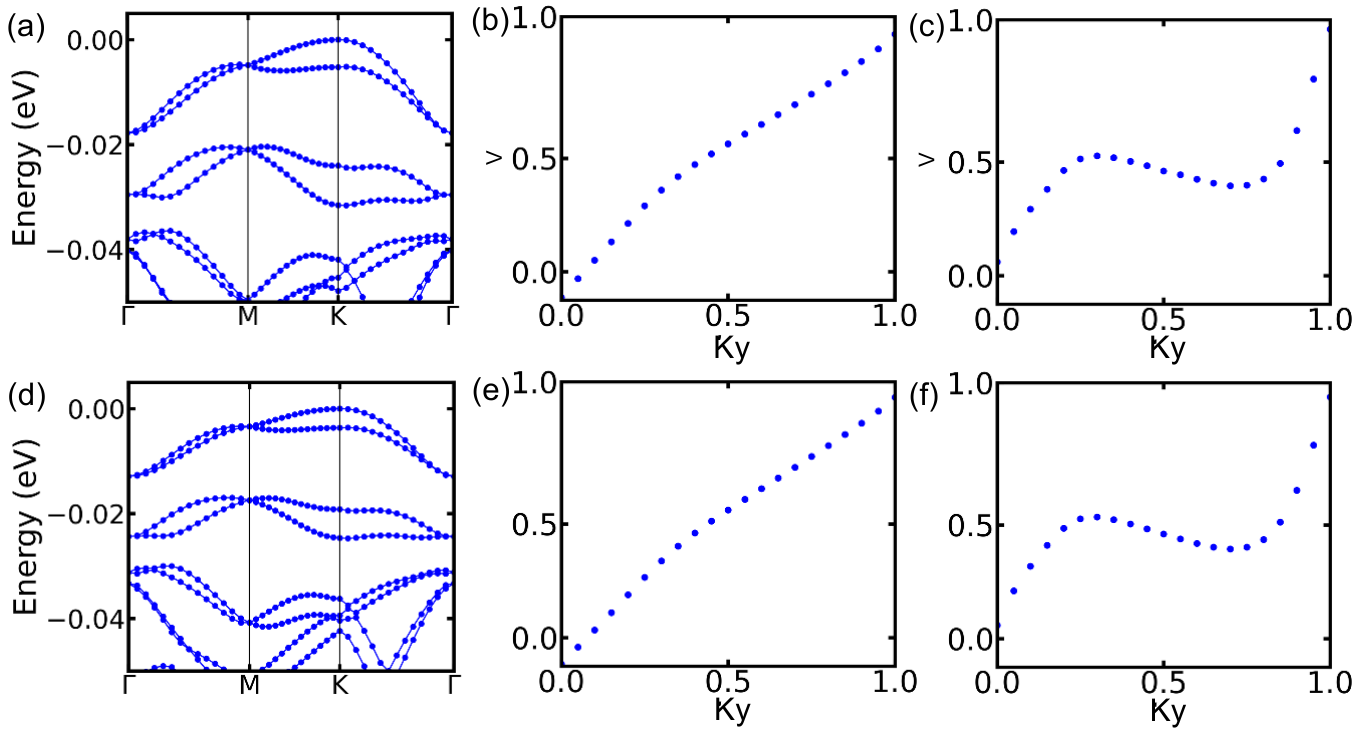}
\caption{
Band structures of strained t-MoTe$_2$ with twist angles of (a) 3.676$^{\circ}$ and (d) 3.305$^{\circ}$.
Panels (b) and (c) show the evolution of the Wannier charge centers for the twist angle of 3.676$^{\circ}$, while panels (e) and (f) correspond to 3.305$^{\circ}$.
The winding of the Wannier charge centers indicates a Chern number of 1 for both the first and second bands, and 1 for the third band.
}\label{strain1}
\end{figure}

During the fabrication of twisted MoTe$_2$, strain is inevitably introduced~\cite{wang2024imaging,thompson2024visualizing}. Therefore, it is essential to investigate its impact on band topology. Accordingly, we applied various perturbations and subsequently computed the topological properties of the system. In particular, we imposed uniaxial stress on one of the two MoTe$_2$ layers before stacking them into a magic-angle structure~\cite{kerelsky2019maximized}. Three different stress configurations were selected. For the first configuration, the twist angle is $3.676^\circ$ with an applied stress of 0.006; for the second, the twist angle is $3.305^\circ$ with a stress of 0.005; and for the third, the twist angle is $2.05^\circ$ with a stress of 0.01. It is noteworthy that, although the applied stresses in all three configurations are less than 0.01, they completely alter the geometric configuration of the material, leading to a total loss of the system's $C_3$ symmetry. In these configurations, the angle between the lattice vectors $\bm{a}_1$ and $\bm{a}_2$ becomes $114^\circ$, $115^\circ$, and $103^\circ$, respectively. Nevertheless, several bands near the Fermi level continue to exhibit nontrivial topological properties. As shown in Fig.~\ref{strain1}, for the two configurations with larger twist angles, the Chern numbers of the two isolated bands remain 1 and 1, similar to the unstressed $3.89^\circ$ structure. This indicates that stress does not change the band topology at larger twist angles.

\section{Symmetry Transformations of $\Omega_{xy}^S$ in the t-MoTe$_2$ (P321) System}
In twisted MoTe\(_2\) with space group P321 (No.\,150),
the relevant symmetry operations include:
\begin{equation}
C_{3z}, \quad 2_{100}, \quad 2_{010}, \quad 2_{110}, \quad \text{and}\quad T.
\end{equation}
We focus on how these symmetry elements transform the \(\Omega_{xy}\) component of the Berry curvature,
where \(\mathbf{k} = (k_x, k_y, k_z)\).
\begin{enumerate}
\item \textbf{Threefold rotation about the \(z\)-axis, \(C_{3z}\):}
\begin{equation}
C_{3z}:\quad \Omega_{xy}(\mathbf{k}) \;\longrightarrow\;
\Omega_{xy}\bigl(R_{3z}(\mathbf{k})\bigr),
\end{equation}
where \(R_{3z}\) rotates the in-plane momentum by \(120^\circ\).
Because this is a proper rotation, \(\Omega_{xy}\) does not change sign
but is evaluated at the rotated wavevector.
\item \textbf{Twofold rotations about in-plane axes, \(2_{100},\,2_{010},\,2_{110}\):}\\
Each of these is a \(180^\circ\) rotation about a vector lying in the crystal plane.
As an example, under \(2_{100}\),
\begin{equation}
k_x \to k_x,\quad k_y \to -k_y,\quad k_z \to -k_z,
\end{equation}
and the antisymmetric tensor component \(\Omega_{xy}\) changes sign:
\begin{equation}
2_{100}:\quad \Omega_{xy}(k_x, k_y, k_z)
\;\longrightarrow\;
-\,\Omega_{xy}\bigl(k_x, -k_y, -k_z\bigr).
\end{equation}
Note that, $2_{100}$ symmetry also flips the spin channel, and give the $J_x$ a minus sign.
\item \textbf{Time-reversal symmetry, \(T\):}\\
Under time reversal,
\begin{equation}
T:\quad \mathbf{k} \,\longrightarrow\, -\,\mathbf{k},
\quad\quad
\Omega_{xy}(\mathbf{k}) \,\longrightarrow\, -\,\Omega_{xy}\bigl(-\mathbf{k}\bigr).
\end{equation}
Time-reversal thus reverses the sign of \(\Omega_{xy}\) and simultaneously inverts~\(\mathbf{k}\).
As before, time-reversal symmetry flip the spin channel, and give the $J_x$ a minus sign.
\end{enumerate}
Overall, we can use all the five symmetries to simplify the calculation.

\section{Two-Stage Dense Matrix GPU Diagonalization}
The first stage is to reduce the original matrix A to a band form matrix B. Given a dense Hermitian matrix $A$, the first step is to reduce it to a band matrix $B$ using a series of Householder transformations. To do that, we compute orthogonal matrices $ Q_1, Q_2, \dots, Q_k $ such that
\begin{equation}
B = Q_k^T \cdots Q_2^T Q_1^T A\, Q_1 Q_2 \cdots Q_k,
\end{equation}
where $ B $ has nonzero elements only in a band of width $ w $ around its main diagonal (with $ w \ll n $). Each Householder reflector $ Q_i $ is defined by
\begin{equation}
Q_i = I - \tau_i v_i v_i^T,
\end{equation}
with $ v_i $ being the Householder vector and $\tau_i$ being the scaling factor
\begin{equation}
\tau_i = \frac{2}{v_i^T v_i}.
\end{equation}
Then, we apply $Q_i$ to matrix $A$,
\begin{equation}
A' = Q_i^T A\, Q_i = A - \tau_i \left( v_i (v_i^T A) + (A v_i) v_i^T \right) + \tau_i^2 (v_i^T A v_i) v_i v_i^T.
\end{equation}

Once we get the band form of $B$, the next step is to further reduce it to a tridiagonal matrix $T$. A symmetric tridiagonal matrix has the structure
\begin{equation}
T = \begin{pmatrix}
\alpha_1 & \beta_1 & 0      & \cdots & 0 \\
\beta_1  & \alpha_2 & \beta_2 & \cdots & 0 \\
0        & \beta_2  & \alpha_3 & \ddots & \vdots \\
\vdots   & \vdots   & \ddots  & \ddots & \beta_{n-1} \\
0        & 0        & \cdots  & \beta_{n-1} & \alpha_n \\
\end{pmatrix}.
\end{equation}
Householder transformations are again used to eliminate the remaining off-band elements. In other words, we obtain
\begin{equation}
T = Q_B^T B\, Q_B,
\end{equation}
where $ Q_B $ is the product of the Householder reflectors used in this stage.
The eigenvalue problem for the tridiagonal matrix is transformed into
\begin{equation}
T y = \lambda y,
\end{equation}
which can be solved very efficiently using methods such as the divide-and-conquer algorithm or QR methods. Because the tridiagonal eigenvalue problem involves only \( O(n^2) \) operations and has a much simpler structure, the two-stage method overall results in a significant performance gain, particularly when implemented on GPUs.

\clearpage

\clearpage

\clearpage

\clearpage

\clearpage

\clearpage

\end{document}